\documentclass[prl,aps,superscriptaddress,footinbib,twocolumn]{revtex4-2}
\usepackage[latin9]{inputenc}
\usepackage{color}
\usepackage{amsmath}
\usepackage{amssymb}
\usepackage{stmaryrd}
\usepackage{graphicx}

\usepackage[ruled]{algorithm2e}

\newcommand{\mcol}{\multicolumn{2}{c}}
\newcommand{\cfill}{\cellcolor[HTML]{c0e0cc}}
\newcommand{\cfillo}{\cellcolor[HTML]{f6deba}}
\usepackage[table,xcdraw]{xcolor}

\usepackage[unicode=true,
 bookmarks=true,bookmarksnumbered=false,bookmarksopen=false,
 breaklinks=false,pdfborder={0 0 1},backref=false,colorlinks=true]
 {hyperref}
\hypersetup{
 linkcolor=magenta, urlcolor=blue, citecolor=blue, pdfstartview={FitH}, hyperfootnotes=false, unicode=true}

\makeatletter

\usepackage{amsfonts}\usepackage{tabularx}\usepackage{dcolumn}\usepackage{bm}\usepackage{graphicx}\usepackage{epstopdf}
\usepackage{times}
\hypersetup{urlcolor=blue}

\def\ket#1{\left|#1\right\rangle}
\def\bra#1{\left\langle#1\right|}

\makeatother

\begin{document}
\title{Observation of a symmetry-protected topological time crystal with superconducting
qubits}

\affiliation{Interdisciplinary Center for Quantum Information, State Key Laboratory  of Modern Optical Instrumentation,
and Zhejiang Province Key Laboratory of Quantum Technology and Device,\\
Department of Physics, Zhejiang University, Hangzhou 310027, China\\
$^2$ Center for Quantum Information, IIIS, Tsinghua University, Beijing 100084, China\\
$^3$ Alibaba-Zhejiang University Joint Research Institute of Frontier
Technologies, Hangzhou 310027, China\\
$^4$ Shanghai Qi Zhi Institute, 41th Floor, AI Tower, No. 701 Yunjin Road, Xuhui District, Shanghai 200232, China\\
$^5$ Hangzhou Global Scientific and Technological Innovation Center, Zhejiang University, Hangzhou 311215, China}

\author{Xu Zhang$^{1, *}$}
\author{Wenjie Jiang$^{2, *}$}
\author{Jinfeng Deng$^{1, *}$}
\author{Ke Wang$^{1}$}
\author{Jiachen Chen$^{1}$}
\author{Pengfei Zhang$^{1}$}
\author{Wenhui Ren$^{1}$}
\author{Hang Dong$^{1}$}
\author{Shibo Xu$^{1}$}
\author{Yu Gao$^{1}$}
\author{Feitong Jin$^{1}$}
\author{Xuhao Zhu$^{1}$}
\author{Qiujiang Guo$^{5, 3}$}
\author{Hekang Li$^{1, 3}$}
\author{Chao Song$^{1, 3}$}
\author{Zhen Wang$^{1, 3, \dagger}$}
\author{Dong-Ling Deng$^{2, 4, \ddagger}$}
\author{H. Wang$^{1, 3, 5}$}

\begin{abstract}
We report the observation of a symmetry-protected topological
time crystal, which is implemented with an array of programmable superconducting
qubits. Unlike the time crystals reported in previous experiments,
where spontaneous breaking of the discrete time translational symmetry
occurs for local observables throughout the whole system, the topological
time crystal observed in our experiment breaks the time translational
symmetry only at the boundaries and has trivial dynamics in the bulk.
More concretely, we observe robust long-lived temporal correlations
and sub-harmonic temporal response for the edge spins up to  40 driving
cycles. We demonstrate that the sub-harmonic response is independent of whether the initial states are random product states or symmetry-protected topological states, and experimentally map out the phase boundary between the time crystalline and thermal phases. 
Our work paves
the way to exploring peculiar non-equilibrium phases of matter emerged
from the interplay between topology and localization as well as periodic
driving, with current noisy intermediate-scale quantum processors.
\end{abstract}

\maketitle
\begin{figure*}
\includegraphics[width=0.98\textwidth]{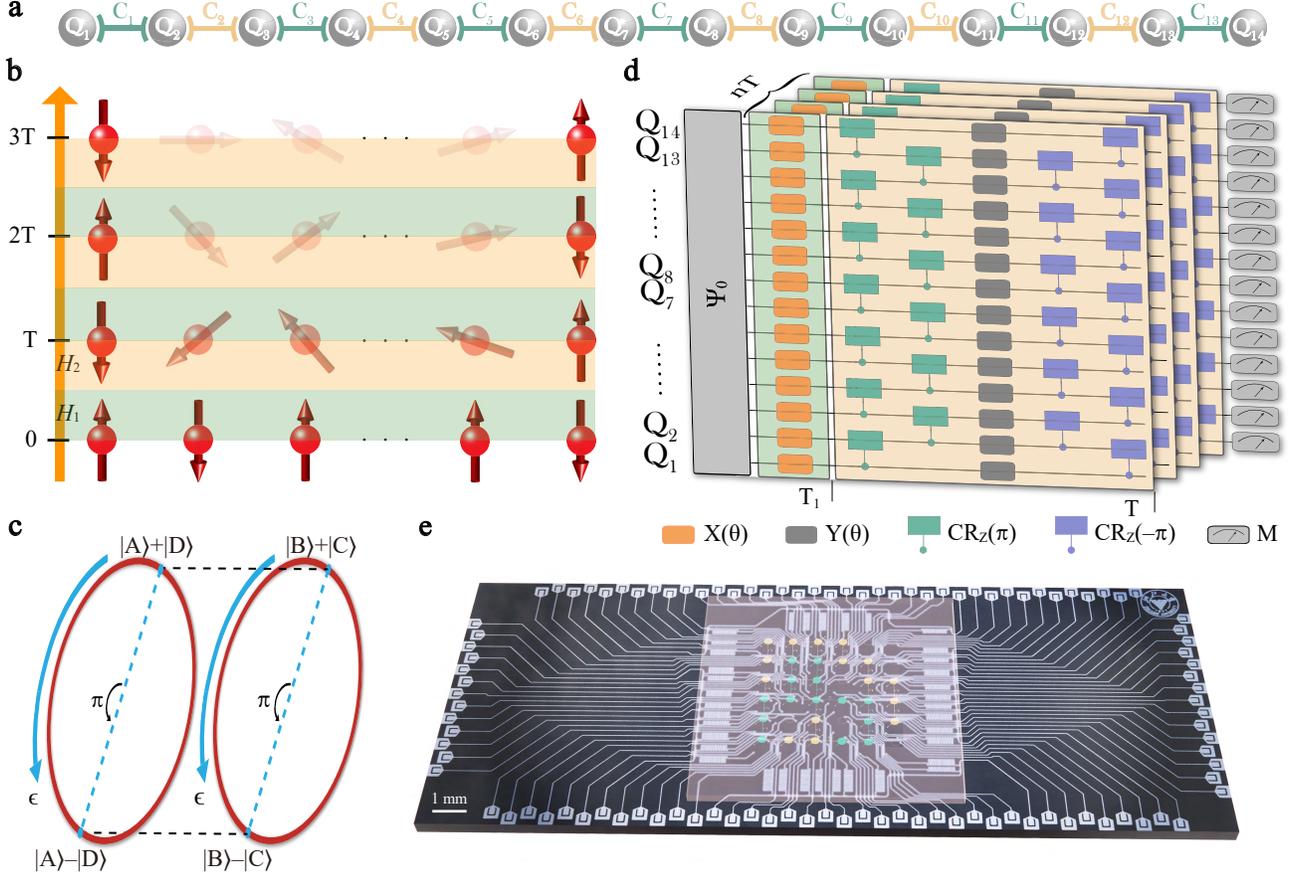}

\caption{Symmetry protected topological time crystal and schematics of the experimental setup.
\textbf{a}, Fourteen qubits used in our experiment are coupled to their neighbors with capacitive couplers. \textbf{b}, A chain of spins are periodically driven with the stroboscopic Floquet Hamiltonian $H(t)$, giving rise to a SPT time crystal characterized by spontaneous time translational symmetry breaking at the
boundaries. \textbf{c}, The quasienergy spectrum ($\epsilon$) of the Floquet unitary,
which is the time evolution operator over one period. For the SPT
time crystal, every eigenstate is two-fold degenerate and has a cousin
separated by quasienergy $\pi$. \textbf{d}, A schematic illustration of
the experimental circuits used to implement the time dynamics governed
by the Floquet Hamiltonian $H(t)$. We randomly sample the Hamiltonians
and prepare the initial states onto random product states or random SPT states.
After running a sequence of quantum gates, we measure
the local magnetization for each qubit or stabilizers at discrete time points. 
\textbf{e}, Illustration of the quantum processor, with the 14 qubits highlighted in green.}
\label{Figure-SPTTCandExpSetup}
\end{figure*}

Quantum many-body systems away from equilibrium host a rich
variety of exotic phenomena that are forbidden by equilibrium thermodynamics.
A prominent example concerns time crystals \cite{Wilczek2012Quantum,Else2016Floquet,Yao2017Discrete,Khemani2016Phase,Sacha2017Time,Else2020Discrete,Yao2018Time,Khemani2019Brief},
where time translational symmetry is spontaneously broken in periodically
driven systems. Pioneering experiments have observed signatures of
time crystalline phases with trapped ions \cite{Zhang2017Observation,Kyprianidis2021Observation},
spins in nitrogen-vacancy centers \cite{Choi2017Observation,O2020Signatures,Randall2021Observation},
ultracold atoms \cite{Smits2018Observation,Autti2018Observation},
solid spin ensembles \cite{Rovny2018Observation,Pal2018Temporal},
and superconducting qubits \cite{Mi2021Observation,Ying2021Floquet,Xu2021Realizing}. Here, we report experimental observation of signatures for a symmetry-protected topological (SPT) time crystal with superconducting qubits (see Fig. \ref{Figure-SPTTCandExpSetup} for an illustration).

In general, SPT states are characterized by non-trivial
edge states that are confined near the boundaries and protected by
certain symmetries \cite{Senthil2015Symmetry,Chiu2016Classification}.
In a clean system without disorder, these edge states in general only
occurs for the ground states of the system with a bulk energy gap.
On raising temperature, they are typically destroyed by mobile thermal
excitations. However, adding strong disorder can make the system many-body
localized (MBL)\cite{Nandkishore2015many, Xu2018MBL}, allowing for a sharply defined
topological phase and stable edge states even at infinite temperature
\cite{Huse2013Localization,Bahri2015Localization}. Strikingly, the topological phase and corresponding edge states can even survive external drives, as long as the driving frequency is large enough so that the localization persists \cite{Harper2020Topology}.

The interplay between symmetry, topology, localization, and periodic driving gives rise to various peculiar phases of matter that exist only out of equilibrium \cite{Harper2020Topology}. Understanding and categorizing these unconventional phases  poses a notorious scientific challenge. On the theoretical side,  topological classifications of periodically driven (Floquet) systems with \cite{Keyserlingk2016PhaseI,Else2016Classification,Potter2016Classification} and without \cite{Roy2017Periodic} interactions have already been obtained through a range of mathematical techniques (such as group cohomology), revealing a number of mysterious phases with no equilibrium counterparts \cite{Harper2020Topology}. Yet, we still lack powerful analytical tools or numerical algorithms to thoroughly address these phases and their transitions to other ones. On the experimental side, observations of signatures for time crystals \cite{Wilczek2012Quantum,Else2016Floquet,Yao2017Discrete,Khemani2016Phase,Sacha2017Time,Else2020Discrete,Yao2018Time,Khemani2019Brief}, which is a paradigmatic example of exotic phases beyond equilibrium \cite{Watanabe2015Absence}, have been reported in a wide range of systems \cite{Zhang2017Observation,Kyprianidis2021Observation,Choi2017Observation,O2020Signatures,Randall2021Observation,Choi2017Observation,O2020Signatures,Randall2021Observation,Smits2018Observation,Autti2018Observation,Rovny2018Observation,Pal2018Temporal,Mi2021Observation,Ying2021Floquet,Xu2021Realizing}. However, none of these experiments encompasses topology as a key ingredient and the realization of a topological time crystal \cite{Giergiel2019Topological,Wahl2021Topologically}, which demands delicate concurrence of topology and localization as well as periodic driving, still remains a notable experimental challenge so far.

In this paper, we report the observation of a SPT time crystal with an array of programmable superconducting qubits. With high controllability and long coherence time, we successfully implement the dynamics of the system under a prototypical Floquet Hamiltonian, which we theoretically predict to exhibit a SPT time crystalline phase. We measure the temporal correlations and the local spin magnetizations, and demonstrate that both of them show a subharmonic response at the boundaries but not in the bulk of the chain. The subharmonic response, which manifests the spontaneous breaking of discrete time-translational sysmmetry carried by the Floquet Hamiltonian, maintains for an extended parameter region and is robust to various experimental imperfections, independent of the initial states. In addition, the SPT time crystal is further explored from the perspectives of entanglement dynamics, entanglement spectrum, and the dynamics of stabilizers that underlies its topological nature. Through measuring the variance of the subharmonic peak height in the Fourier spectrum, we also experimentally map out the phase boundary between the time crystalline and thermal phases. The observed SPT time crystal in our experiment differs drastically from all other conventional ones with trivial topology,  
which opens a door for harnessing this exotic phase of matter for practical quantum information processing  \cite{Preskill2018Quantum}.

\vspace{.5cm}
\noindent\textbf{\large{}Model Hamiltonian}{\large\par}

\noindent We consider an one-dimensional (1D) spin-$\frac{1}{2}$ chain governed by the following Floquet Hamiltonian (see Fig. \ref{Figure-SPTTCandExpSetup}\textbf{b}):
\begin{eqnarray}
H(t) & = & \begin{cases}
H_{1}, & \text{for } 0\leq t<T_{1}\\
H_{2}, & \text{for } T_{1}\leq t<T
\end{cases}\label{eq:FloquetHam} \\
H_{1} & \equiv & (\frac{\pi}{2}-\delta)\sum_{k}\hat{\sigma}_{k}^{x}\label{eq:TQCH1},\\
H_{2} & \equiv & -\sum_{k}[J_{k}\hat{\sigma}_{k-1}^{z}\hat{\sigma}_{k}^{x}\hat{\sigma}_{k+1}^{z}+V_{k}\hat{\sigma}_{k}^{x}\hat{\sigma}_{k+1}^{x}+h_{k}\hat{\sigma}_{k}^{x}],\label{eq:TQCH2}
\end{eqnarray}
where $\hat{\sigma}^{x,z}_k$ is the Pauli matrix acting on the $k$-th spin; 
$J_{k}$, $V_{k}$, and $h_{k}$ are random parameters drawn independently
from uniform distributions over $[J-\Delta_{J},J+\Delta_{J}]$, $[V-\Delta_{V},V+\Delta_{V}],$
and $[h-\Delta_{h},h+\Delta_{h}]$, respectively. For simplicity,
we fix $T=2T_{1}=2$ throughout this paper. We note that $H(t)$ has a $Z_{2}\times Z_{2}$
symmetry. For a suitable parameter region, it has been shown that $H_2$ can be in a MBL phase, where topological edge states can survive as coherent degree of freedom at arbitrarily high energies \cite{Bahri2015Localization}. The localization and edge states carry over to the case of periodic driving with the Hamiltonian $H(t)$, giving rise to a  SPT time crystalline phase (see Supplementary Materials I. B). 

The Floquet unitary that fully characterizes the SPT time crystal reads $U_F=U_2U_1$, where $U_{1}=e^{-iH_{1}}$ and $U_2=e^{-iH_2}$ are the unitary operators generated by the Hamiltonian $H_1$ and $H_2$, respectively. To understand why time translational symmetry only breaks at the boundary but not in the bulk, we consider the idealized cluster limit ($V_k=h_k=0$) and set $\delta=0$. We suppose that the system is initially prepared to a random product state in the computational basis, and we use the dynamics of local magnetization as a diagnosis. In this simple scenario, the non-trivial structure of the cluster states (eigenstates of $U_2$) gives rise to edge modes that behave as free spins. At each driving period, the unitary operator $U_1$ basically flips all spins. As a result, the edge spins are reversed after one period and return to their initial configuration after two, leading to the doubling response periodicity for the local magnetization at the boundaries. For spins in the bulk, however, the unitary operator $U_2$ plays a role and evolves the random product state to a state with vanishing magnetization, resulting in no double periodicity or symmetry breaking. A closer look at the quasienergy specturm of $U_F$ reveals that its every eigenstate is two-fold degenerate with a long-range spatial order and has a cousin eigenstate separated by quasienergy $\pi$ (see Fig. \ref{Figure-SPTTCandExpSetup}\textbf{c}). This is essential for the robustness of the subharmonic response of the edge spins against local perturbations.

\begin{figure*}[tbp]
\centering
\includegraphics[width=1.0\linewidth]{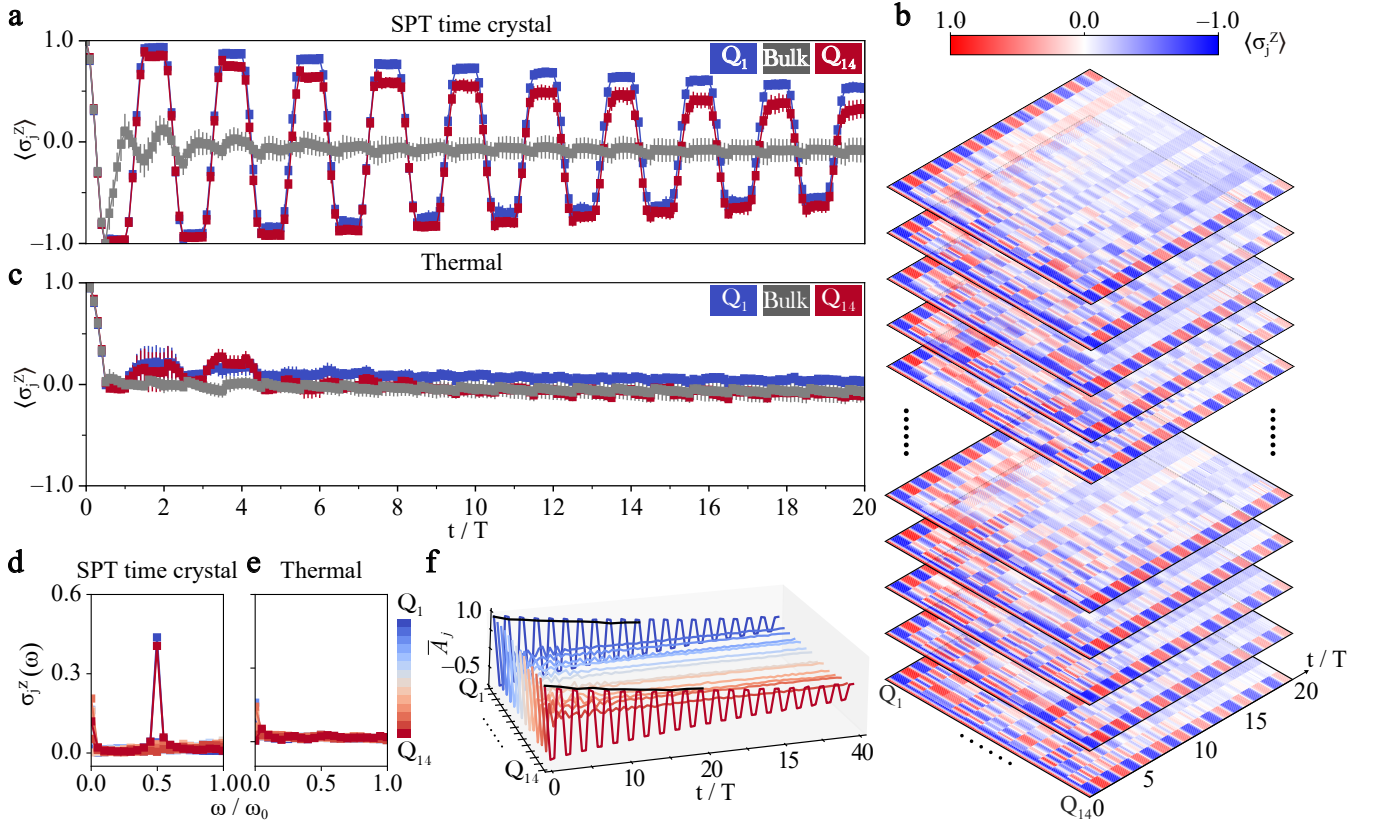}
\caption{
Observation of a SPT time crystal. \textbf{a},
Time evolution of disorder-averaged local magnetizations deep in the SPT
time crystal region ($J=\Delta_J=1$, $V=h=\Delta_V=\Delta_h=0$, and $\delta=0.01$). The initial
states are $\ket{0}^{\otimes{L}}$ and data shown are
averaged over 20 random disorder instances. While the bulk magnetization
decays quickly to zero, the edge spins oscillate with a stable subharmonic
response up to 20 cycles. \textbf{b}, The evolution dynamics of local magnetizations for different random instances. Here, each layer corresponds to a specific random instance.  \textbf{c}, Magnetization dynamics deep in the thermal region ($J=\Delta_J=1$, $V=h=\Delta_V=\Delta_h=0$, and $\delta=0.8$). \textbf{d},
Fourier transform of experimentally
measured $\langle\sigma_{j}^{z}(t)\rangle$ in the time crystal region. The edge spins lock to
the subharmonic frequency, which is in sharp contrast to that for
bulk spins. \textbf{e}, Fourier spectra of $\langle\sigma_{j}^{z}(t)\rangle$ in the thermal region. No robust subharmonic frequency peak appears for both edge and bulk spins in this case. \textbf{f}, Time-dependence of the autocorrelator $\overline{A}_j=\overline{\langle \Psi_0|\sigma_j^z(t)\sigma_j^z(0)|\Psi_0\rangle}$ up to 40 cycles, obtained from averaging over $20$ random instances deep in the SPT time crystalline phase, with the initial states prepared at random product states $\ket{\{0, 1\}}^{\otimes{L}}$. The black solid lines show the results of ``echo'' circuits for two boundary qubits.}
\label{fig:ExpTimeEv}
\end{figure*}

\vspace{.5cm}
\noindent\textbf{\large{}Experimental setup}{\large\par}

\noindent Our experiment is performed on a flip-chip superconducting quantum processor designed to encapsulate a square array of $6 \times 6$ transmon qubits with adjustable nearest-neighbor couplings (Fig.~\ref{Figure-SPTTCandExpSetup}\textbf{e}), on which a chain of up to $L=14$ qubits, denoted as $Q_1$ through $Q_{L}$, that alternate with $L-1$ couplers, denoted as $C_1$ through $C_{L-1}$, are selected to observe the SPT time crystal (Fig.~\ref{Figure-SPTTCandExpSetup}\textbf{a}). All $L$ qubits can be individually tuned in frequency with flux biases, excited by microwaves, and measured using on-chip readout resonators; all couplers are also of transmon type with the characteristic transition frequencies higher than those of the qubits, which can be controlled with flux biases to tune the effective nearest-neighbor couplings. During an experimental sequence (Fig.~\ref{Figure-SPTTCandExpSetup}\textbf{d}), we first initialize each qubit, $Q_j$, in $|0\rangle$ at its idle frequency $\omega_{j}$, following which we alternate the single-qubit gates at $\omega_{j}$ with the two-qubit controlled-$\pi$ (CZ) gates realized by biasing $Q_j$ and its neighboring qubit to the pairwise frequencies listed in $\left[\omega_{j}^{\text{A (B)}}, \omega_{j+1}^{\text{A (B)}}\right]$ for a fixed interaction time (see Supplementary Materials III. C). Meanwhile, each coupler is dynamically switched between two frequencies~\cite{Yan2018Tunable,Arute2019QuantumSupremacy,Yan2020CZ,Collodo2020CZ,Sung2021CZ,Wu2021StrongAdvantage}: one is to turn off the effective coupling where the neighboring two qubits can be initialized and operated with single-qubit gates; the other one is to turn on the nearest-neighbor coupling to around 11~MHz for a CZ gate. After $n$ number of layers of the alternating single- and two-qubit gates, we finally tune all qubits to their respective $\omega_{j}^{\text{m}}$ for simultaneous quantum-state measurement. Qubit energy relaxation times measured around $\omega_{j}$ are in the range of 11 to 37 $\mu$s. More characteristic qubit parameters, including the above mentioned frequencies, anharmonicities, and readout fidelities, can be found in Supplementary Materials Tab.~S1.

We explore a quantum digital simulation scheme to implement the dynamics of the system under the driven Hamiltonian $H(t)$. More specifically, we decompose the evolution operators to the experimentally feasible single-qubit gates [X$(\theta)$, Y$(\theta)$, and Z$(\theta)$], and the two-qubit gates [CR$_z(\pm\pi)$], where X$(\theta)$, Y$(\theta)$, and Z$(\theta)$ are rotations around $x$-, $y$-, and $z$-axis by the angle $\theta$, respectively, and CR$_z(\pm\pi)$ are the $z$-axis rotations of the target qubit by $\pm\pi$ conditioned on the state of the control qubit (see Fig.~\ref{Figure-SPTTCandExpSetup}\textbf{d} and Supplementary Materials III. A for the ansatz that generating the gate sequences). X$(\theta)$ and Y$(\theta)$ are realized by applying 50 ns-long microwave pulses with full width half maximum  of 25 ns, whose quadrature correction terms
are optimized to minimize state leakages to higher levels\cite{Song201710q}. Simultaneous randomized benchmarkings indicate that the single-qubit gates used in this experiment have reasonably high fidelities, averaging above 0.99 (see Supplementary Materials Tab.~S1). Z$(\theta)$ is realized using the virtual-Z gate, which encodes the information $\theta$ in the rotation axes of all subsequent gates\cite{Mckay2017EfficientZ}, and is combined with CZ to assemble CR$_z(\pm\pi)$. Here, we adopt the strategy reported elsewhere~\cite{Foxen2020FSim,Sung2021CZ} to realize the CZ gate, i.e., diabaticly tune the coupler frequency while keeping $|11\rangle$ and $|02\rangle$ (or $|20\rangle$) for the subspace of the two neighboring qubits in near resonance. The 40 ns-long CZ gate for a pair of neighboring qubits can be individually optimized to be around 0.99 in fidelity as calibrated by interleaved randomized benchmarking; when simultaneously running the CZ gates for multiple pairs of neighboring qubits as required in the experimental sequence, the averaged CZ gate fidelities can be around 0.985 as obtained by simultaneous randomized benchmarkings (see Supplementary Materials Tab.~S1).

\vspace{.5cm}
\noindent \textbf{\large{}Symmetry breaking at boundaries}{\large\par}

\noindent The characteristic signature of a SPT time crystal is the breaking of the discrete time-translational symmetry at the boundaries of the chain but not in the bulk. This can be manifested by the persistent oscillation with period $2T$ of local magnetizations at the boundaries. In Fig. \ref{fig:ExpTimeEv}, we plot the time evolution of disorder-averaged local magnetizations $\sigma_j^{z}(t)$ for different phase regions.  From Fig. \ref{fig:ExpTimeEv}\textbf{a}, it is evident that in the SPT time crystal region, the disorder-averaged magnetizations at the two ends of the chain, namely $\sigma_1^{z}(t)$ and $\sigma_{L}^{z}(t)$, oscillate with a $2T$ periodicity, up to over $20$ driving cycles. In stark contrast, the magnetizations at the bulk of the chain ($\sigma_j^{z}(t)$ with $2\leq j \leq L-1$) decay quickly to zero and do not show periodic-doubled oscillations. This unconventional behavior is independent of disorder average. Even for a single random disorder instance, the magnetizations show similar dynamical features, as shown in Fig. \ref{fig:ExpTimeEv}\textbf{b}.  The distinction between the  dynamics of boundary and bulk magnetizations can also be clearly seen by examining $\sigma_j^{z}(t)$ in the frequency domain. As shown in Fig. \ref{fig:ExpTimeEv}\textbf{d}, 
the edge spins lock to the subharmonic frequency of the drive period $\omega/\omega_0=1/2$ whereas the bulk spins show no such peak. We stress that the subharmonic response for the edge spins obtained in our experiment is notably robust to various perturbations and experimental imperfections (see Supplementary Materials I. B for a more in-depth discussion).  For 
comparison, we also experimentally measure the dynamics of the magnetizations in the thermal region. Our results are shown in Fig. \ref{fig:ExpTimeEv}\textbf{c} and Fig. \ref{fig:ExpTimeEv}\textbf{e}, where we see that the magnetizations for both the edge and bulk spins decay quickly to zero and no subharmonic response appears at all.

The spontaneous breaking of the discrete time translational symmetry at the boundaries can also be detected by the autocorrelators defined as $\overline{A}_j=\overline{\langle \Psi_0|\sigma_j^z(t)\sigma_j^z(0)|\Psi_0\rangle}$. Our experimental result for autocorrelators up to 40 driving cycles is plotted in Fig. \ref{fig:ExpTimeEv}\textbf{f}, showing the breaking of time translational symmetry at the boundaries but not in the bulk again. We mention that in the SPT time crystal region the local magnetizations for the edge spins exhibit a gradually decaying envelope, which is attributed to either the external circuit errors or slow internal thermalization. To distinguish these two mechanisms, we carry out an additional experiment about the ``echo'' circuit $U_{\text{echo}}\equiv (U_F^{\dagger})^tU_F^t$, whose deviation from the identity operation measures the effect of circuit errors \cite{Mi2021Observation}. The square root at the output of $U_{\text{echo}}$ (black solid lines shown in Fig.~\ref{fig:ExpTimeEv}\textbf{f}) fits well with the decaying envelop of $U_{F}$, indicating that the decay of the envelope is due to circuit errors rather than thermalization, which assures that the system is indeed in the MBL region.

\begin{figure}
\includegraphics[width=0.49\textwidth]{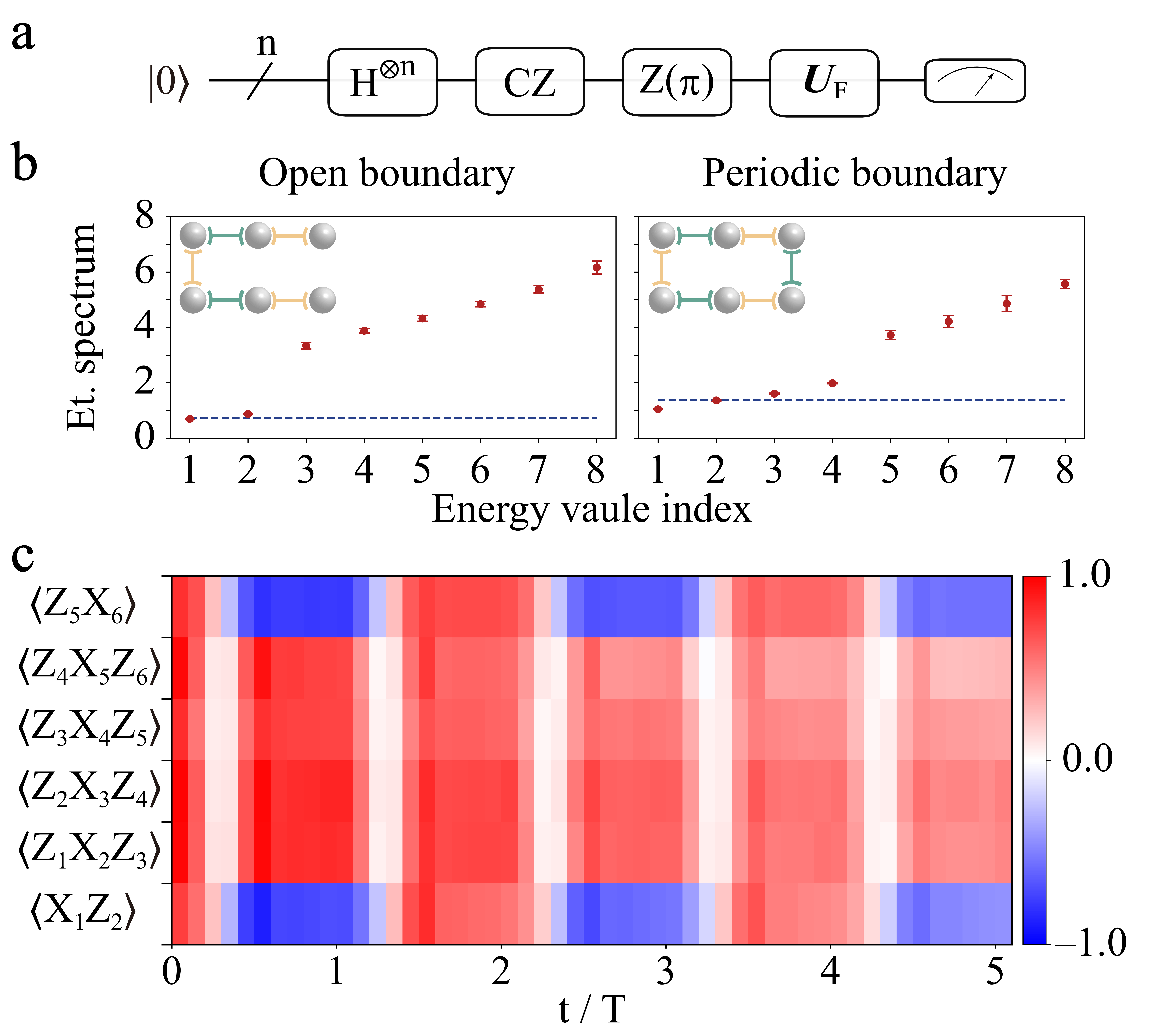}
\caption{Dynamics of stabilizers with random initial SPT states. \textbf{a}, Schematic of the experimental circuit for preparing random SPT states. To prepare the system into the ground state of the cluster Hamiltonian $H'_2$, we apply a Hardmard gate (H) on each qubit and then run CZ gates in parallel for all neighboring qubit pairs in two steps. Then we act Z$(\pi)$ on some random sites to create excitations and transfer the ground state to a highly excited eigenstate of $H'_2$. This procedure enables the preparation of random SPT states at high energy. We then evolve these states with the Floquet 
Hamiltonian $H(t)$ to study the dynamics of stabilizers.  \textbf{b}, Entanglement spectra of random SPT states prepared in our experiment, with both open and periodical boundary conditions. The dashed lines indicate the lowest degenerate entanglement energies obtained by simulation in error-free case. The two- and four-fold degeneracy at low entanglement energy is a characterizing feature for the topological nature of these states. \textbf{c}, The time dependence of stabilizers in the SPT time crystal region, averaged over $20$ random circuit instances.}
\label{fig:SPTstates}
\end{figure}

\vspace{.5cm}
\noindent\textbf{\large{}Localization-protected topological states}{\large\par}

\noindent In the above discussion, the initial states are random product states. To establish SPT time crystal, additional experiments on other initial states and other local observables are advisable. 
In this section, we show that the stabilizers in the bulk do not break the discrete time translational symmetry, but at the boundaries do. To understand this, we consider the idealized cluster and spin-flip limit, i.e., $V_k=h_k=0$ and $\delta=0$. In this limit, $H_2$ reduces to a summation of stabilizers: $H_2=-\sum_{k=2}^{L-1}J_{k}S_k$ with $S_k\equiv \hat{\sigma}_{k-1}^{z}\hat{\sigma}_{k}^{x}\hat{\sigma}_{k+1}^{z}$. To break the degeneracy of $H_2$,
we consider adding two boundary terms $J_1S_1$ ($S_1\equiv \hat{\sigma}_1^x\hat{\sigma}_2^z$) and $J_LS_L$ ($S_L\equiv \hat{\sigma}_L^x\hat{\sigma}_{L-1}^z$), which are commuting with all bulk stabilizers, into the Hamiltonian $H_2$ and form a new cluster Hamiltonian  $H_2'=H_2+J_1S_1+J_LS_L$. We note that the eigenstates of $H_2'$ are also eigenstates of $H_2$ with degeneracy split by the boundary terms.  We choose the initial states to be random eigenstates of $H_2'$ and evolve the system with the Floquet Hamiltonian $H(t)$. We measure the time-dependence of local stabilizers.   

In Fig. \ref{fig:SPTstates}\textbf{a}, we show a sketch of the quantum circuit used in our experiment to prepare the desired random eigenstates of $H_2'$. To manifest the topological nature of these eigenstates, we study their entanglement spectra \cite{Li2008Entanglement}, which are widely used as a crucial diagnostic for universal topological properties of quantum phases \cite{Li2008Entanglement,Swingle2012Geometric,Fidkowski2010Entanglement,Alba2012Boundary}.  In our experiment, we prepare random eigenstates of $H_2$ with both open and periodic boundary conditions and  obtain the reduced density matrices of half of the system $\rho_{\text{half}}$ through quantum state tomography. Figure \ref{fig:SPTstates}\textbf{b} displays the entanglement spectra ($-\ln v$ where $v$ stores the eigenvalues of $\rho_{\text{half}}$) of two experimentally prepared eigenstates for the open and periodic boundary condition, respectively.  From this figure, a clear two-fold degeneracy for the low entanglement energy Schmidt states is obtained for the open boundary condition, which corresponds to the spin half degrees of freedom released at the bipartitioning boundary. Whereas, for the periodic boundary condition, the spectrum is four-fold degenerate, corresponding to two effectively decoupled spins at the two boundaries of the bipartition. The degeneracy of the entanglement spectra and its dependence on boundary conditions marks a characteristic feature of the SPT state prepared in our experiment. 
We remark that the degeneracy disappears at high entanglement energy. This is due to the finite size effects and experimental imperfections. 

In Fig. \ref{fig:SPTstates}\textbf{c}, we plot the time-dependence of local stabilizers. In the SPT time crystal region, we observe that the stabilizers at the boundaries oscillate with a $2T$ periodicity, indicating again a spontaneous breaking of the discrete time translational symmetry at the boundaries.  In the bulk, the stabilizers oscillate with a $T$ periodicity and are synchronized with the driving frequency, thus no symmetry breaking occurs. This is in sharp contrast to the dynamics of bulk magnetizations, which decays rapidly to zero and exhibits no oscillation. In fact, in the SPT time crystal region the system is MBL and there exist a set of local integrals of motion, which are the ``dressed'' version of the stabilizers with exponentially small tails \cite{Bahri2015Localization}.   
The persistent oscillations of the bulk stabilizers observed in our experiment originates from these local integrals of motion and is a reflection of the fact that the system is indeed in a MBL phase.

\begin{figure}
\includegraphics[width=0.49 \textwidth]{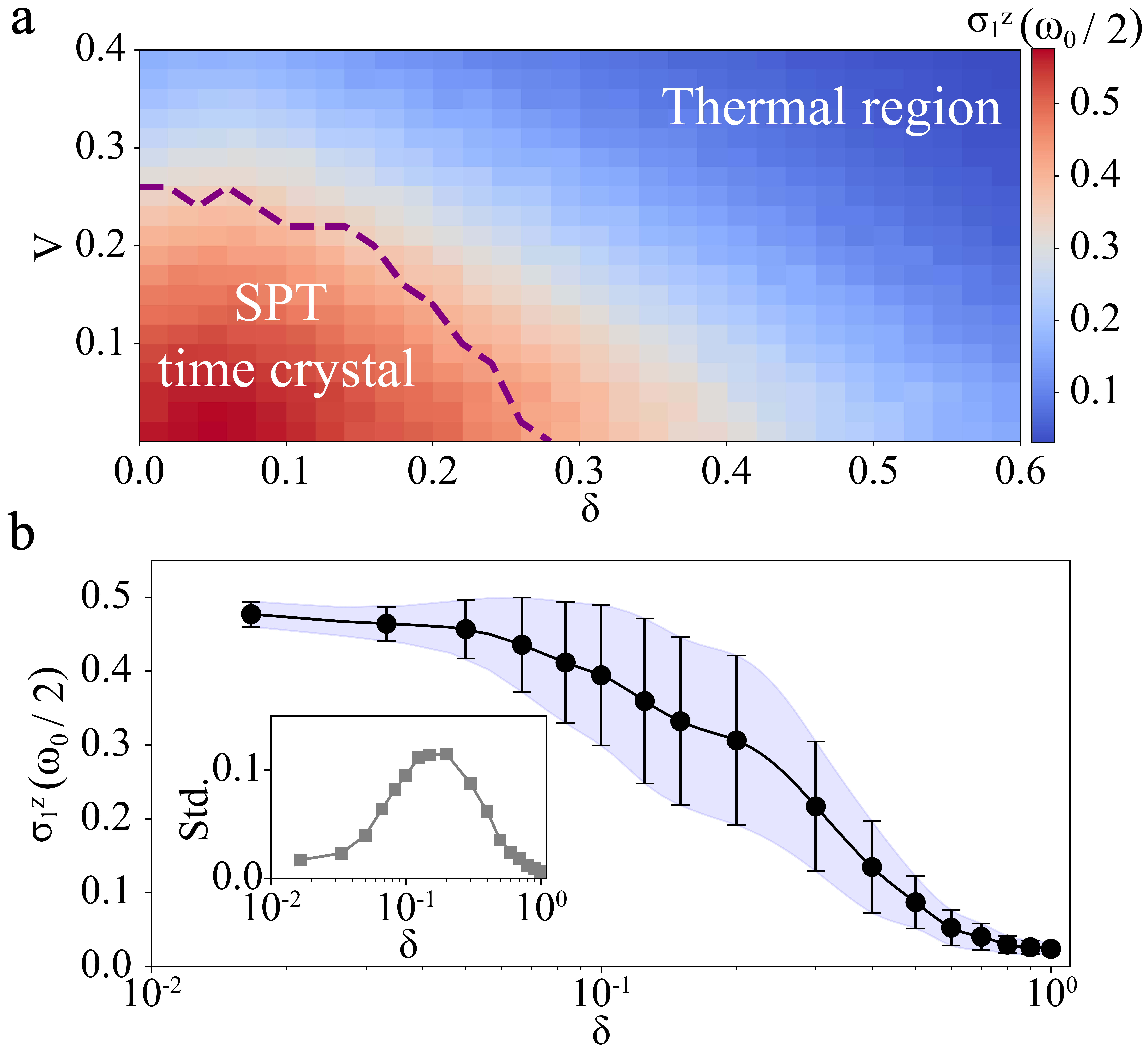}

\caption{Phase diagram and detection of phase transition.  \textbf{a}, The numerical $\delta$-$V$ phase diagram obtained by examining the central subharmonic peak height for the edge spins in the Fourier spectrum, averaged over 1000 disorder instances. The dashed line corresponds to the maximal height variances for varying $V$ with each fixed $\delta$ point, which gives a rough estimation of the phase boundary. Here, the parameters are chosen as $L=8$, $J=\Delta_J=1$, and $h=\Delta_h=\Delta_V=0.05$.   \textbf{b}, Experimental result of the subharmonic peak height as a function of the drive perturbation $\delta$ with fixed $h=V=\Delta_h=\Delta_V=0$ and $J=\Delta_{J}=1$, averaged over 20 disorder instances uniformly sampling from the interval $[J-\Delta_{J},J+\Delta_{J}]$ for 14 qubits, 
with the shadow outlining  the standard deviation. Inset, the standard deviation of the central peak height as a function of $\delta$, which signals a phase transition from the SPT time crystalline phase to a thermal phase.
\label{fig:PhaseTransition}}
\end{figure}

\vspace{.5cm}
\noindent \textbf{\large{}Phase transition}{\large\par}
\noindent We now turn to the phase transition between the SPT time crystalline phase and the trivial thermal phase. For simplicity and concreteness, we fix other parameters and vary the drive perturbation $\delta$ and strength of the interactions $V$. For the chosen parameter regions, theoretically the system would exhibit a SPT time crystalline phase with small $\delta$ and $V$. Whereas, with increasing $\delta$ and $V$ the strong interaction diminishes localization and eventually thermalizes the system. At some critical values of $\delta$ and $V$, a transition between these two phases occurs. In Fig. \ref{fig:PhaseTransition}\textbf{a}, we plot the $\delta$-$V$ phase diagram obtained from numerical simulations, where the phases boundary, although not very sharp due to the finite-size effect, can be located and visualized approximately.

To experimentally examine this phase transition, we further fix the interaction strength  $V=0$. 
We probe the transition point by measuring the variance of the subharmonic spectral peak height, i.e., Fourier spectrum of $\langle\sigma_{1}^{z}\rangle$ at $\omega~=~\omega_0/2$ for the boundary spin.  Figure \ref{fig:PhaseTransition}\textbf{b} shows the subharmonic peak height as a function of the drive perturbation $\delta$. With small $\delta$, the system is in the SPT time crystalline phase and the peak height remains at a notable value about $0.5$. As we increase $\delta$ to a large value, the crystal `melts' and peak height vanishes. This confirms the theoretical analysis above.  The largest variance of the peak height corresponds to the phase transition point. The inset of Fig. \ref{fig:PhaseTransition}\textbf{b} shows the measured standard deviation as a function of $\delta$, indicating a phase transition point around  $\delta\approx 0.2$.

\vspace{.5cm}
\noindent \textbf{\large{}Conclusion and outlook}{\large\par}
\noindent In summary, we have experimentally observed signatures of a SPT time crystal with a programmable superconducting quantum processor. In contrast to previously reported conventional time crystals, for our observed SPT time crystal the discrete time translational symmetry only breaks at the boundries but not in the bulk. We measured the persistent oscillations of edge spins with a subharmonic frequency and experimentally demonstrated that the SPT time crystal phase is robust against to perturbations in the drive and imperfections in the experiment. In addition, we also demonstrated that subharmonic response of the boundary observables is independent of the initial states.

The controllability and scalability of the superconducting platform demonstrated in our experiment open up several new avenues for future fundamental studies and potential applications. In particular, it would be interesting and important to explore other exotic non-equilibrium phases beyond classical simulability, with the superconducting or other  platforms. 
In practice, the observed SPT time crystal may have applications in some quantum information processing tasks, such as quantum metrology or implementing a robust quantum memory with topological protection.

\vspace{.5cm}
\noindent\textbf{\large{}Acknowledgement}{\large\par}

The device was fabricated at the Micro-Nano Fabrication
Center of Zhejiang University.  We acknowledge 
the support of the National Natural Science
Foundation of China (Grants No. 11725419, U20A2076,
92065204, and 12075128), the National Basic Research
Program of China (Grants No. 2017YFA0304300), 
the Zhejiang Province Key Research
and Development Program (Grant No. 2020C01019), 
and the Key-Area Research and Development Program
of Guangdong Province (Grant No. 2020B0303030001). 
D.-L. D. also acknowledges additional support from the Shanghai Qi Zhi Institute.

\noindent{* These authors contributed equally to this work.\\
$^\dagger$ 2010wangzhen@zju.edu.cn\\
$^\ddagger$ dldeng@tsinghua.edu.cn}

\clearpage

\setcounter{secnumdepth}{3}

\makeatletter
\setcounter{figure}{0}
\setcounter{equation}{0}
\renewcommand{\thefigure}{S\@arabic\c@figure}
\renewcommand \theequation{S\@arabic\c@equation}
\renewcommand \thetable{S\@arabic\c@table}

\begin{center} 
	{\large \bf Supplementary Materials: Observation of a symmetry-protected topological time crystal with superconducting qubits}
\end{center}

\section{Theoretical Understanding}

\subsection{Introduction to the conventional Floquet time crystals}
In order to obtain a better intuitive understanding of the  symmetry-protected topological (SPT) time crystal, we first introduce the basic concepts of the traditional Floquet time crystals and give a prototypical model as a concrete example.

Spontaneous symmetry-breaking is an important concept in modern physics. This occurs when the steady states of a physical system do not respect the symmetries of the Hamiltonian governing this system. An important example manifests spontaneous symmetry-breaking is the ordinary crystals, which break the continuous spatial translational symmetry. More precisely, in a crystal, the system is not invariant under continuous translation operators, which is respected by the system's Hamiltonian. Analogously, systems that spontaneously break the time translational symmetry are named as time crystals \cite{Shapere2012Classical,Wilczek2012Quantum}. Although there is a no-go theory for the continuous time crystals \citet{Bruno2013Impossibility}, Floquet time crystals manifests themselves in many physical systems. There are two equivalent definitions of Floquet time crystals in Ref \cite{Else2016Floquet}, which characterize this concept from the perspective of the expectation value of an operator and from the perspective of the eigenstates of the Floquet evolution unitary respectively. The first definition states that time translational symmetry breaking occurs if for every state $\ket{\psi(t)}$ at arbitrary time $t$, there exists an operator $O$ satisfying that $\bra{\psi(t+T)}O\ket{\psi(t+T)}\neq\bra{\psi(t)}O\ket{\psi(t)}$, where $\ket{\psi(t+T)}=U_F(T)\ket{\psi(t)}$ with $U_F(T)$ being the Floquet evolution unitary of one period $T$. This definition implies how to observe time crystals experimentally and is used in our paper. The second definition states that the time translational symmetry breaking occurs if all eigenstates of the Floquet evolution unitary are long-range correlated. This concept is used in our theoretical analysis.

To be more concrete, we introduce the following prototypical time-dependent Hamiltonian previously studied in Ref \cite{Else2016Floquet,Yao2017Discrete} as an example of Floquet time crystals:
\begin{equation}\nonumber
H_{F}(t)=\left\{ \begin{array}{rl}
H_{1} & =\pi/2\sum_{k}\hat\sigma_{k}^{x},\ \ \ \ \ \ \ \ \ \ 0<t<T_{1}\\
H_{2} & =\sum_{k}J_{k}\hat\sigma_{k}^{z}\hat\sigma_{k+1}^{z}+h_{k}^{z}\hat\sigma_{k}^{z}\ \ \ \ \ T_{1}<t<T
\end{array}\right.
\end{equation}
where $J_{k}$ and $h_{k}^{z}$ are uniformly
chosen from $J_{k}\in[J/2,3J/2]$, $h_{k}^{z}\in[0,h^{z}]$. For simplicity, we choose $T=2T_{1}=2$. The
Floquet evolution operator in one period can be written as $U_{F}=\exp(-iH_{2})\exp(-i\pi/2\sum_{k}\hat\sigma_{k}^{x})$.

We consider the eigenstates of $H_{2}$, which are the product states
polarized in $z$ direction $\ket{\Theta}=\ket{\{s_{k}\}}$ with $s_k=\pm1$, and these states can be prepared in experiments. As evolution unitary in the first time interval has the effect of flipping all spins, the state $\ket{\Theta}$ is related to another state $\ket{-\Theta}=\ket{\{-s_{k}\}}$, which is also an eigenstate of $H_2$. As for the time evolution in the second time interval, operating unitary $e^{-iH_{2}}$ on the two
states, we will have $e^{-iH_{2}}\ket{\Theta}=\exp[-i(E^{+}({\Theta})+E^{-}{(\Theta)})]\ket{\Theta}$,
$e^{-iH_{2}}\ket{-\Theta}=\exp[-i(E^{+}({\Theta})-E^{-}{({\Theta})})]\ket{-{\Theta}}$,
where $E^{+}(\Theta)$ and $E^{-}(\Theta)$ are the expectation value
of the first and the second part of $H_{2}$ for the state $\ket{\Theta}$. Therefore, under Floquet operator $U_{F}$, we have relations: 
\begin{equation}
U_{F}\ket{\Theta}=\exp[-i(E^{+}({\Theta})-E^{-}({\Theta}))]\ket{-\Theta}
\end{equation}
\begin{equation}
U_{F}\ket{-\Theta}=\exp[-i(E^{+}({\Theta})+E^{-}({\Theta}))]\ket{\Theta}
\end{equation}
From the above equations, we see the Floquet evolution unitary $U_{F}$ mixes the two states $\ket{\Theta}$
and $\ket{-\Theta}$. In the subspace formed by $\ket{\pm\Theta}$,
$U_{F}$ has the matrix form:
\[
U_{F}=\begin{bmatrix}0 & e^{-i(E^{+}(\Theta)-E^{-}(\Theta))}\\
e^{-i(E^{+}(\Theta)+E^{-}(\Theta))} & 0
\end{bmatrix}.
\]
Diagonalize the matrix, we can get eigenvalues $\pm\exp(-iE^{+}(\Theta))$,
and eigenstates $\ket{\Theta}\pm\exp{[iE^{-}(\Theta)]}\ket{-\Theta}$. Obviously, the eigenstates of $U_{F}$ are paired cat states with long-range correlations. Thus, in the localized region, this model satisfies the second definition of a Floquet time crystal in Ref. \cite{Else2016Floquet} and discrete time translational symmetry breaking could occur in such a system. Besides, as the Floquet operator has eigenvalue $\pm\exp(-iE^{+}(\Theta))$, if we diagonalize the effective Hamiltonian of the Floquet operator, we can get two eigenvalues with quasi-energy difference $\pi$. This is just the $\pi$-spin glass phase introduced in \cite{Khemani2016Phase}.

\subsection{Our model: the SPT time crystal}
Different from the model mentioned above, our SPT time crystal only breaks the discrete time translational symmetry at the boundaries. Precisely speaking, our model manifests the subharmonic response at frequency $2\pi/2T$ only at the edges, but not in the bulk of the system. Here $T$ is the period of the Floquet driving. Now, let us give more theoretical analysis of this model.

\subsubsection{Localized and SPT quantum states}

Our SPT time crystal has two distinct governing Hamiltonians in different time intervals as showing in the main text. In the first time interval, this governing Hamiltonian is the sum of one-body Pauli operators on different sites. And in the second time interval, the governing Hamiltonian includes interaction among neighboring sites, which introduces the subtle many-body properties in this system. Let us begin with the static Hamiltonian $H_2$ \cite{Bahri2015Localization}: 
\begin{equation}
    H_{2} =-\sum_{k}\left[J_{k} \hat{\sigma}_{k-1}^{z} \hat{\sigma}_{k}^{x} \hat{\sigma}_{k+1}^{z}+V_{k} \hat{\sigma}_{k}^{x} \hat{\sigma}_{k+1}^{x}+h_{k} \hat{\sigma}_{k}^{x}\right],\nonumber
\end{equation}
where the parameters are chosen as explained in the main text. This Hamiltonian has a $Z_{2}\times Z_{2}$ symmetry, corresponding
to $\hat\sigma_{k}^{z,y}\longrightarrow-\hat\sigma_{k}^{z,y}$ independently
on even or odd numbered sites, i.e. $[H_2,\prod_k\hat\sigma_{2k}^x]=0$ and $[H_2,\prod_k\hat\sigma_{2k+1}^x]=0$. The principle terms in this Hamiltonian are those three-body operators $S_k=\hat{\sigma}_{k-1}^{z} \hat{\sigma}_{k}^{x} \hat{\sigma}_{k+1}^{z}$, which are commuting with each others, i.e. $[S_k, S_l]=0$. They are called stabilizers. 

In the extreme case $V_k=h_k=0$, the eigenstates of this Hamiltonian are the mutual eigenstates of all stabilizers, which are named as cluster states. They are SPT states with $Z_2\times Z_2$ symmetry. The SPT phase manifests itself in the open boundary case: there is one effective free spin at each end of the chain. The topological property of those states is encapsulated by the string-order parameter:
\[O_{\mathrm{st}}(l, j)=\langle\hat\sigma_{l}^{z} \hat\sigma_{l+1}^{y}(\prod_{k=l+2}^{j-2} \hat\sigma_{k}^{x}) \hat\sigma_{j-1}^{y} \hat\sigma_{j}^{z}\rangle,\]
which takes random values $O_{\mathrm{st}}(l, j)=\pm1$ for different eigenstates and different disorder realizations. Thus, we can define a non-local analogue of the Edwards-Anderson glass-order parameter to characterize the SPT time crystal phase: $O_{\text{sg}}=[\![O^2(l,j)]\!]$, where $[\![]\!]$ denotes an average over sites, states and random realizations. Furthermore, the entanglement spectra of the SPT states are degenerate. This degeneracy can serve as another manifestation of the topological property. Besides, in this case, all energy levels are exactly four-fold degenerate. The corresponding degenerate eigenstates can be divided into four groups: $\{\ket{A_{k}}=\ket{\uparrow...\uparrow}\}$, $\{\ket{B_{k}}=\ket{\downarrow...\uparrow}\}$, $\{\ket{C_{k}}=\ket{\uparrow...\downarrow}\}$, $\{\ket{D_{k}}=\ket{\downarrow...\downarrow}\}$. Here we are working
in the $\hat\sigma_{z}$ basis for simplicity, and the two arrows represent the effective boundary spins and the $...$ denotes the bulk spins. They are related by $\prod\hat\sigma_{\text{odd}}^{x}\ket{A_{k}}=\ket{B_{k}}$, $\prod\hat\sigma_{\text{even}}^{x}\ket{A_{k}}=\ket{C_{k}}$, $\prod\hat\sigma_{\text{all}}^{x}\ket{A_{k}}=\ket{D_{k}}$.

When $V_k,\ h_k\neq 0$, the two body terms and the one-body terms make the eigenstates of this Hamiltonian departing from the cluster states. However, if we keep this Hamiltonian deep in the topological phase (the region we are interested in): $J_k\gg V_k, h_k$, we can also interpret this model from a many-body localized (MBL) perspective. Different from the Hamiltonian with strictly localized stabilizers being integrals of motion, in this region, the system obtains another set of mutually commuting quasi-local integrals of motion. Each of them has a finite interaction length, and thus, is not strictly local. Similarly, in open boundary condition, there exists a quasi-local effective free spin at each edge, which contains the bulk components decaying exponentially with the distance. In this case, the string-order parameter and the degeneracy of the entanglement spectra can also manifest the topological property of the corresponding eigenstates. Moreover, while the energy spectra stop to be exactly four-fold degenerate for a finite system size, it is still nearly four-fold degenerate and the corresponding eigenstates can be divided into four groups as mentioned above.

\begin{figure}
    \includegraphics[width=0.48\textwidth]{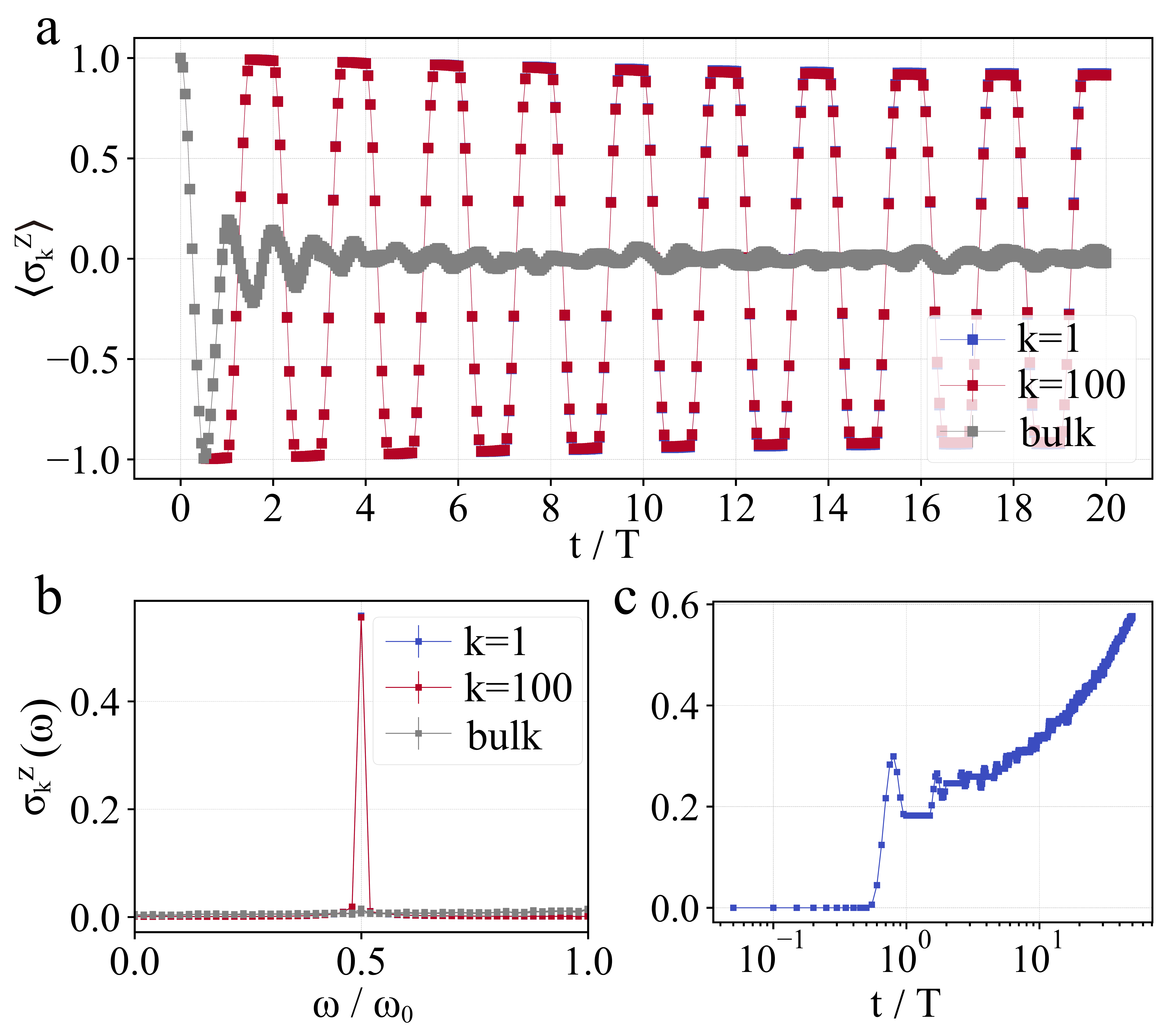}
    \caption{The evolution of the SPT time crystal with system size $L=100$, computed using the time-evolving block decimation methods. Results showed here are averaged over 1000 random realizations, with parameters $J=\Delta J=1$, $h=\Delta h=V=\Delta V=\delta=0.05$. \textbf{a}. Time evolution of disorder-averaged local observables. For the edge spins, it is clear that $\langle\sigma^z_1$ and $\langle\sigma^z_{100}\rangle$ displays persistent oscillations with a $2T$ periodicity, manifesting the breaking of discrete time-translational symmetry. In stark contrast, in the bulk region $\langle\sigma^z_k\rangle$ decays rapidly to zero and no symmetry breaking is observed. This is the
    defining feature of the SPT time crystals: time-translational symmetry only breaks at the boundary, not in the bulk. \textbf{b}. Fourier spectra of $\langle\sigma_{k}\rangle$. We find that $\sigma_{1}(\omega)$ has a peak at $\omega/\omega_{0}=1/2$, where $\omega_{0}=2\pi/T$ is the driving frequency. We mention that this peak is robust and rigidly locked at $\omega_{0}/2$, a manifestation of the robustness of the SPT time crystals. For the bulk spins, there is no such peak, consistent with no symmetry breaking in the bulk. \textbf{c}. Logarithmic entanglement entropy growth. In the SPT time crystal region, the system is many-body localized. We thus expect a logarithmic entanglement growth, which is showing in this figure. We note that the entanglement has a initial quick rise till time $Jt\sim1$, corresponds to expansion of wave packets to a size of the order of the localization length.\label{fig:dmrg}}
\end{figure}

\subsubsection{The emergence of the SPT time crystal}

With those understandings of the static Hamiltonian $H_2$, now let us consider the Floquet case, wherein we periodically drives the above SPT Hamiltonian as discussed in the paper:
\begin{eqnarray}
    H(t) & = & \begin{cases}
    H_{1}, & 0\leq t<T_{1}\\
    H_{2}, & T_{1}\leq t<T
    \end{cases}\nonumber\label{eq:FloquetHam-1},\\
    H_{1} & \equiv & (\lambda-\delta)\sum_{k}\hat{\sigma}_{k}^{x},\nonumber\label{eq:TQCH1-1}\\
    H_{2} & \equiv & -\sum_{k}[J_{k}\hat{\sigma}_{k-1}^{z}\hat{\sigma}_{k}^{x}\hat{\sigma}_{k+1}^{z}+V_{k}\hat{\sigma}_{k}^{x}\hat{\sigma}_{k+1}^{x}+h_{k}\hat{\sigma}_{k}^{x}].\nonumber\label{eq:TQCH2-1}
\end{eqnarray}

Let us begin with the perfect case, where $\lambda=\pi/2,\ \delta=0$ and $V_k,\ h_k=0$. So the energy spectrum of $H_2$ is perfectly four-fold degenerate. The eigenstates can be divided into four groups, i.e. $\{\ket{A_{k}}=\ket{\uparrow...\uparrow}\}$, $\{\ket{B_{k}}=\ket{\downarrow...\uparrow}\}$, $\{\ket{C_{k}}=\ket{\uparrow...\downarrow}\}$, $\{\ket{D_{k}}=\ket{\downarrow...\downarrow}\}$ with $E_{A_k}=E_{B_k}=E_{C_k}=E_{D_k}$. There exists a local effective free spin at each boundary. Since the effect of $U_1=e^{-i\pi/2\sum_k\hat\sigma_k^x}$ is perfectly flipping the spins at all sites, we can obtain the following relations of the Floquet operator $U_F=\exp \left(-i H_{2}\right) \exp \left(-i \pi / 2 \sum_{k} \hat{\sigma}_{k}^{x}\right)$:
\begin{equation}
    U_{F}\ket{A_{k}}=\exp(-iH_{2})\ket{D_{k}}=\exp(-iE_{D_{k}})\ket{D_{k}},
\end{equation}
\begin{equation}
    U_{F}\ket{B_{k}}=\exp(-iH_{2})\ket{C_{k}}=\exp(-iE_{C_{k}})\ket{C_{k}},
\end{equation}
\begin{equation}
    U_{F}\ket{C_{k}}=\exp(-iH_{2})\ket{B_{k}}=\exp(-iE_{B_{k}})\ket{B_{k}},
\end{equation}
\begin{equation}
    U_{F}\ket{D_{k}}=\exp(-iH_{2})\ket{A_{k}}=\exp(-iE_{A_{k}})\ket{A_{k}}.
\end{equation}
From this, we see that $U_{F}$ mixes the states $\ket{A_{k}}$ with
$\ket{D_{k}}$, and mixes the states $\ket{B_{k}}$ with $\ket{C_{k}}$.
Writing in the subspace of $\ket{A_{k}}$ and $\ket{D_{k}}$, $U_{F}$
has the matrix form:
\[
\begin{bmatrix}0 & \exp(-iE_{D_{k}})\\
\exp(-iE_{A_{k}}) & 0
\end{bmatrix}.
\]
In the subspace of $\ket{B_{k}}$ and $\ket{C_{k}}$, $U_{F}$ has
the matrix form:
\[
\begin{bmatrix}0 & \exp(-iE_{C_{k}})\\
\exp(-iE_{B_{k}}) & 0
\end{bmatrix}.
\]
Therefore, in the subspace formed by $\ket{A_{k}}$, $\ket{B_{k}}$, $\ket{C_{k}}$, and $\ket{D_{k}}$, $U_{F}$ has eigenvalues $\pm\exp[-i(E_{A_{k}}+E_{D_{k}})/2]$ and $\pm\exp[-i(E_{B_{k}}+E_{C_{k}})/2]$. Thus, the floquet effective Hamiltonian has eigen-energies: $({E_{A_{k}}+E_{D_{k}}})/2$, $({E_{B_{k}}+E_{C_{k}}})/2$, $({E_{A_{k}}+E_{D_{k}}})/2+\pi$, $({E_{B_{k}}+E_{C_{k}}})/2+\pi$ $\mod2\pi$. As the energy spectrum is four-fold degenerate: $E_{A_k}=E_{B_k}=E_{C_k}=E_{D_k}$, the Floquet eigen-energies should have this relation: $({E_{A_{k}}+E_{D_{k}}})/2=({E_{B_{k}}+E_{C_{k}}})/2$,\ $({E_{A_{k}}+E_{D_{k}}})/2+\pi=({E_{B_{k}}+E_{C_{k}}})/2+\pi$. Therefore, the original four-fold degeneracy breaks into two-fold degeneracy after the drive. This two-fold degeneracy is the reminiscence of the original topological order. As for the Floquet eigenstates, by diagonizing $U_{F}$ in the two subspaces, we will get that these Floquet eigenstates are cat-like linear combination of topological eigenstates: $\ket{A_{k}}\pm\ket{D_{k}}$, and $\ket{B_{k}}\pm\ket{C_{k}}$. These states are long-range correlated, and thus, the mutual information between the two boundaries is $2\log2$. The long-range correlated property of these eigenstates indicates that discrete time translational symmetry breaking can occur in this system, implying a Floquet time crystal.

When we turn on the two-body terms and the one-body terms, but still constrain the Hamiltonian $H_2$ deep in the topological region: $J_{k}\gg h_{k},V_{k}$, it has four nearly degenerate eigenstates related by the symmetry operations for a finite system size. The effective free spin in each boundary becomes quasi-local. Under the Floquet driving, the nearly four-fold degeneracy breaks into nearly two-fold degeneracy: $\left(E_{A_{k}}+E_{D_{k}}\right) / 2\approx\left(E_{B_{k}}+E_{C_{k}}\right) / 2,\left(E_{A_{k}}+E_{D_{k}}\right) / 2+\pi\approx\left(E_{B_{k}}+E_{C_{k}}\right) / 2+\pi$. Following the similar discussion, we can find that the eigenstates of the Floquet unitary are also cat-like states, and thus, time translational symmetry breaking can occur in this case. This discussion makes it clear that our model satisfies the definition of Floquet time crystals in this case, as long as the governing Hamiltonian in the second time interval is also in the deep topological region. The stability of the SPT time crystal will be discussed more precisely in the later subsection.

\subsubsection{Dynamical properties of the SPT time crystal}

Next, we will consider the evolution of this system, and explicitly manifest the behavior of the SPT time crystal.

Let us start from a product state: $\ket{\psi_{0}}=\ket{\downarrow...\uparrow}$, here $...$ denotes the product states of bulk spins. Because the boundary spins fall into the group of $\{\ket{B_{k}}\}$, we can expand the initial state as $\ket{\psi_{0}}=\sum_{k}b_{k}\ket{B_{k}}$. Under the time evolution of $U_{F}$ for one driving period: 
\begin{align}
U_{F}\ket{\psi_{0}} & =\exp(-iH_{2})\exp(-i\pi/2\sum_{j}\hat\sigma_{j}^{x})\sum_{k}b_{k}\ket{B_{k}}\nonumber \\
 & =\exp(-iH_{2})\sum_{k}b_{k}\ket{C_{k}}\\
 & =\sum_{k}b_{k}\exp(-iE_{C_{k}})\ket{C_{k}},\label{eq:1}
\end{align}
where $\ket{B_{k}}=\ket{\downarrow...\uparrow}$, $\ket{C_{k}}=\ket{\uparrow...\downarrow}$.
So if we measure the edge spins at the starting point, we have $\bra{\psi_{0}}\hat\sigma_{1}^{z}\ket{\psi_{0}}=-1$,
$\bra{\psi_{0}}\hat\sigma_{N}^{z}\ket{\psi_{0}}=1$. After one Floquet
period, the state becomes $\ket{\psi_{1}}=\sum_{k}b_{k}\exp(-iE_{C_{k}})\ket{C_{k}}$.
Because $\ket{C_{k}}$ has definite spin expectation values, we will
get $\bra{\psi_{1}}\hat\sigma_{1}^{z}\ket{\psi_{1}}=1$, $\bra{\psi_{1}}\hat\sigma_{N}^{z}\ket{\psi_{1}}=-1$.
Similarly, after two Floquet periods, the state becomes $\ket{\psi_{2}}=\sum_{k}b_{k}\exp(-iE_{C_{k}}-iE_{B_{k}})\ket{B_{k}}$,
and $\bra{\psi_{2}}\hat\sigma_{1}^{z}\ket{\psi_{2}}=-1$, $\bra{\psi_{2}}\hat\sigma_{N}^{z}\ket{\psi_{2}}=1$.
Thus, we see the edge spins exhibit breaking of the time-translational symmetry. 

As for the bulk spins, assume that one bulk spin $\hat\sigma_{j}^{z}$
has expectation value for the initial product state: $\bra{\psi_{0}}\hat\sigma_{j}^{z}\ket{\psi_{0}}=1$.
Writing $\ket{\psi_{0}}$ in the $\ket{B_{k}}$ basis, we have: 
\begin{equation}
\bra{\psi_{0}}\hat\sigma_{j}^{z}\ket{\psi_{0}}=\sum_{k,k^{'}}b_{k}b_{k}^{'}\bra{B_{k^{'}}}\hat\sigma_{j}^{z}\ket{B_{k}}=1.
\end{equation}
Since the spins of $\ket{C_{k}}$ are opposite to the spins of $\ket{B_{k}}$ at all sites, we immediately have that : 
\begin{equation}
\sum_{k,k^{'}}b_{k}b_{k}^{'}\bra{C_{k^{'}}}\hat\sigma_{j}^{z}\ket{C_{k}}=-1.
\end{equation}
However, the expectation value of $\hat\sigma_{j}^{z}$ in the state $\ket{\psi_{1}}$
should be expressed as: 
\begin{equation}
\bra{\psi_{1}}\hat\sigma_{j}^{z}\ket{\psi_{1}}=\sum_{k,k^{'}}b_{k}b_{k}^{'}\exp(-iE_{C_{k}}+iE_{C_{k^{'}}})\bra{C_{k^{'}}}\hat\sigma_{j}^{z}\ket{C_{k}}.\nonumber
\end{equation}
Comparing the two above equations, we see that because of the extra phase
factor $\exp \left(-i E_{C_{k}}+i E_{C_{k^{\prime}}}\right)$ before each component, the
$\hat\sigma_{j}^{z}$ will not have definite value after the Floquet time evolution and will decay to zero quickly after random averaging. Thus, the bulk spins do not exhibit breaking of the time-translational symmetry. 

The above derivations tell that for our model, with the time evolution, the edge spins exhibit discrete time-transnational symmetry breaking, while the bulk spins relax very fast. Thus, the time translational symmetry breaking only occurs at the boundaries as showing in Fig.~\ref{fig:dmrg}\textbf{a}. We stress the importance of topology here. It protects the edge spins, ensuring the robustness of the edge spins against local perturbations that respect the underlying symmetry.

Besides, for a localized system, the entanglement entropy will grow logarithmically \cite{Bardarson2012Unbounded}.  In the deep SPT time crystal region, the system represented by the static Hamiltonian $H_2$ has a series of quasi-local integrals of motion. So, spins far away from each other can build significant entanglement only after exponentially long evolving time. Thus, under the Floquet time evolution, the entanglement entropy of this system will exhibit a logarithmic growth, which is showing in Fig.~\ref{fig:dmrg}\textbf{b}, and will finally saturate to a value proportional to the system size.

\begin{figure}
    \includegraphics[width=0.48\textwidth]{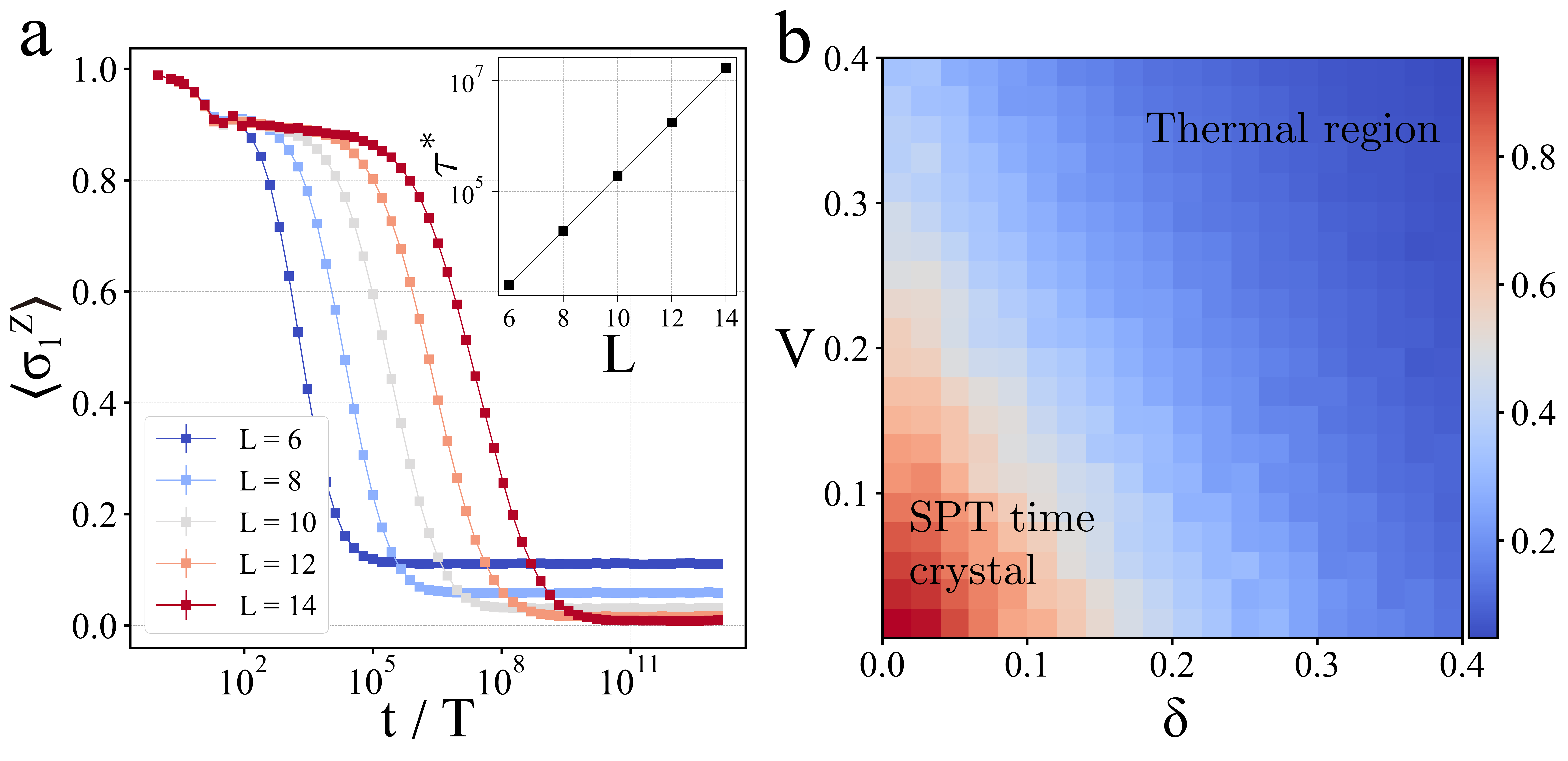}
    \caption{The decay of spins at boundaries and the phase diagram of the SPT crystal. \textbf{a.} The decay of the random realization averaged magnetization for the spin at the edge. Here, the number of the disorder realization ranges from $3\times 10^4$ ($L=6$) to $10^3$ ($L=14$). The omitted parameters are chosen as in Fig.~\ref{fig:dmrg}. We find there is an initial quick decay of the $\langle\sigma_1^z\rangle$, followed by a plateau that extends up to a time diverging exponentially with the system size. The inset shows the exponential scaling of $\tau^*$ with the system size, where $\tau^*$ is the time when the spin at edge decays to $1/2$. \textbf{b.} The phase diagram of the SPT time crystal with respect to the imperfect drive $\delta$ and the average strength of the interactions $V$. Here, we adapt the string order parameter $O_{\text{sg}}$ (averaged over $100$ random realizations) as the indicator. It shows that when the imperfections are not very large, the string order parameter is approximately equal to one, indicating the topological property. The other parameters are chosen as $J=\Delta J=1$, $h=\Delta h=0.01$.\label{fig:life}}
\end{figure}

Furthermore, in the deep topological region, when the system size is finite, the Floquet time evolution can lead the states of the system to the thermal states after long enough evolving time. Actually, the quasi-local effective free spins at the boundaries have exponentially decaying tails, containing the bulk components. When the system size is finite, the tails of the quasi-local effective free spins have an exponentially small overlap, which will relax the two effective free spins. Due to the fact that the overlap is exponentially small with the system size, the lifetime of the time translational symmetry breaking states will exponentially diverge with system size. This phenomenon is demonstrated in our numerical simulation Fig.~\ref{fig:life}\textbf{a}.

\subsubsection{The stability of the SPT time crystal}
The above considerations rely on the fact that during each period,
we perfectly flip spins at all sites. To show that this is indeed a time crystal, we should make sure this phase is stable even for imperfect drive. We follow the similar arguments introduced in Ref. \cite{Else2016Floquet}.

In above subsection, we already showed that in the perfect drive case, the eigenstates of the Floquet evolution operator are cat-like states $\ket{\psi_{\pm}^{AD}}=\ket{A_{k}}\pm\ket{D_{k}}$, and $\ket{\psi_{\pm}^{BC}}=\ket{B_{k}}\pm\ket{C_{k}}$. We say that an effective short-range correlated topological state is that $\bra{\psi}\hat\sigma_{1}\hat\sigma_{N}\ket{\psi}-\bra{\psi}\hat\sigma_{1}\ket{\psi}\bra{\psi}\hat\sigma_{N}\ket{\psi}\rightarrow0$. Obviously, $\ket{A_{k}},\ket{B_{k}},\ket{C_{k}},\ket{D_{k}}$ are short-range correlated topological states, but the Floquet eigenstates $\ket{\psi_{\pm}^{AD}}$ and $\ket{\psi_{\pm}^{BC}}$ are all long-range correlated, with different eigen-energies. Then for any experimentally prepared short-range states (such as product states), they can only be formed by taking superposition of those long-range Floquet eigenstates with different energies. Thus, after one period of Floquet evolution, local observables at the edge will not be invariant and has to break the discrete time translational symmetry. Now we add local perturbations into the system, such as imperfect drives and other local interactions. Since the system is in the MBL phase, a local perturbation will only significantly affect those near sites. Thus, we expect that the long-range correlations in the eigenstates of the Floquet unitary will not disappear. Actually, there exists one quasi-local unitary operator $U$, which relates the perturbed Floquet eigenstates with the un-perturbed Floquet eigenstates. Since $U$ is a quasi-local operator, it cannot destroy the long-range correlations of the un-perturbed Floquet eigenstates. Therefore, time translational symmetry breaking can also occur in the locally perturbed system. To explicitly show that the SPT time crystal is indeed a physical phase, we use the string order parameter $O_{\text{sg}}$ as the indicator to plot the phase diagram with respect to the imperfect drive ($\delta$) and the average strength of the two-body interactions ($V$), which is showing in Fig.~\ref{fig:life}\textbf{b}.

\section{Details of TEBD method}
\begin{figure}[t]
    \includegraphics[width=0.5\textwidth]{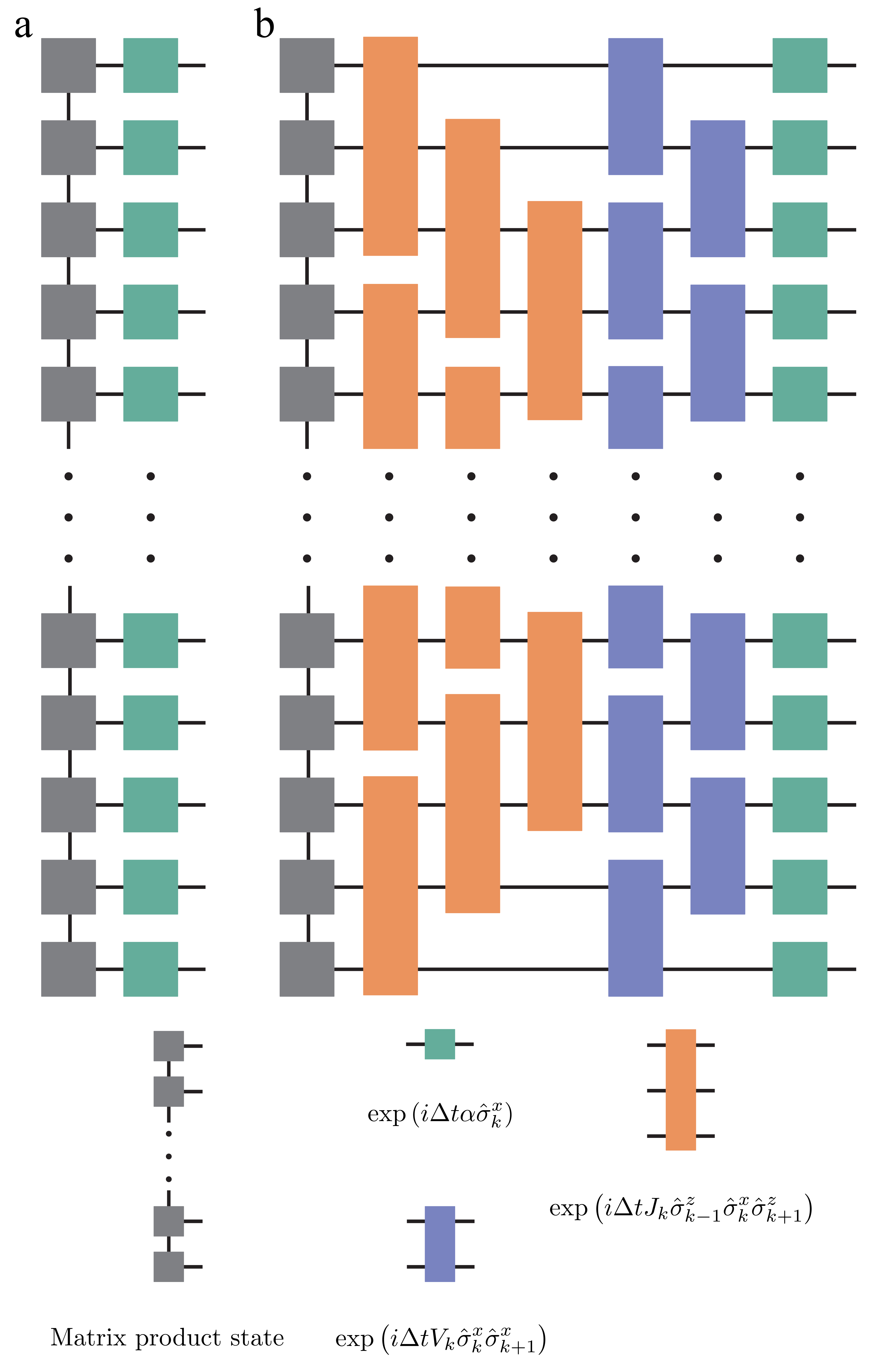}
    \caption{Pictorial illustration of the implementation of the time evolution unitary, where a connected wire between different blocks means contraction of indices. \textbf{a.} Implementation of $U_1(\Delta t)$. The gray blocks represent the current state in MPS form and the green blocks represent the time evolution unitary consisting of one-body operators ($\alpha=-(\pi/2-\delta)$). \textbf{b.} Implementation of $U_2(\Delta t)$. The gray blocks represent the current state in MPS form; the orange blocks represent the time evolution unitary consisting of three-body operators arranged in three groups; the blue blocks represent the time evolution unitary consisting of two-body operators arranged in two groups and the green blocks represent the time evolution unitary consisting of one-body operators ($\alpha=h_k$). They are applied to the the current state layer by layer.}\label{MPO}
\end{figure}
We numerically simulate the time evolution process of the SPT time crystals using the time-evolving block decimation (TEBD) methods. This method was proposed for the time evolution of the matrix product states (MPS) \cite{Vidal2003Efficient,Vidal2004Efficienta} and is variant of the  density matrix renormalization group (DMRG) algorithm \cite{White2004RealTime,Daley2004Timedependent}. At the heart of the TEBD method lies the  Trotter-Suzuki decomposition of the time evolution operator $U(\Delta t)$ of a short-range interaction system in a small time interval $\Delta t$. Usually, we can represent the operator $U(\Delta t)$ in the matrix product operator (MPO) form with small Trotter error, and then repeatedly apply it on the MPS representing current quantum state $\ket{\psi(t)}$ to implement the time evolution.

Our SPT time crystal has two distinct Hamiltonian operators in different time intervals as showing above. For the first time interval, the corresponding Hamiltonian is the sum of one-body operators on different sites. So the evolution operator is a direct product of one-body evolution operators
\begin{equation}
    U_1(t)=e^{-itH_1}=e^{-it(\lambda-\delta)\hat\sigma_1^x}\otimes...\otimes e^{-it(\lambda-\delta)\hat\sigma_k^x}\otimes...,\nonumber
\end{equation}
which can be represented as MPO directly. To obtain the corresponding expectation values of some local observables at different times, we also decompose the time evolution operator of an entire time interval $T_1$ into several small time intervals $\Delta t$. We show the implementation of $U_1(\Delta t)$ in Fig.~\ref{MPO}\textbf{a}.

For the second time interval, the Hamiltonian $H_2$ consists of multiple short-range interaction terms: $H_2=-\sum_k[J_{k} \hat{\sigma}_{k-1}^{z} \hat{\sigma}_{k}^{x} \hat{\sigma}_{k+1}^{z}+V_{k} \hat{\sigma}_{k}^{x} \hat{\sigma}_{k+1}^{x}+h_{k} \hat{\sigma}_{k}^{x}]$. Thus, we can approximate the time evolution operator using Trotter-Suzuki decomposition $U_2(t)\approx \left[U_2(\Delta t)\right]^{t/\Delta t}=\left[e^{-i\Delta tH_2}\right]^{t/\Delta t}$ with $\Delta t\ll t$. To efficiently construct the MPO representation of $U_2(\Delta t)$, we group together terms in $H_2$ which are commuting with each other. For the three-body operators, all of them are stabilizer operators and they are commuting with each other. For the two-body terms, we can group terms without overlap and construct corresponding MPO operators. For one-body terms, all of them are operators on different sites and thus, they are commuting with each other. For simplicity, we denote $A=-\sum_kJ_{k} \hat{\sigma}_{k-1}^{z} \hat{\sigma}_{k}^{x} \hat{\sigma}_{k+1}^{z}, B=-\sum_kV_{2k} \hat{\sigma}_{2k}^{x} \hat{\sigma}_{2k+1}^{x}, C=-\sum_kV_{2k+1} \hat{\sigma}_{2k+1}^{x} \hat{\sigma}_{2k+2}^{x}, D=-\sum_kh_{k} \hat{\sigma}_{k}^{x}$ and have approximation as following
\begin{align}\nonumber
    U_2(\Delta t)=&e^{-i\Delta t (A+B+C+D)}\\\nonumber
    =&e^{-i\Delta t(D+C)}e^{-i\Delta t(B+A)}e^{-i\Delta t^2[D+C,B+A]}\\\nonumber
    &+\mathcal{O}(\Delta t^3)\\\nonumber
    =&e^{-i\Delta tD}e^{-i\Delta tC}e^{-i\Delta tB}e^{-i\Delta tA}+\mathcal{O}(\Delta t^2).\nonumber
\end{align}
Thus, the evolution operator of $H_2$ of time interval $t$ has approximation 
\begin{align}\nonumber
    U(t)&\approx \left[U_2(\Delta t)\right]^{t/\Delta t}\\\nonumber
    &=\left(e^{-i\Delta tD}e^{-i\Delta tC}e^{-i\Delta tB}e^{-i\Delta tA}\right)^{t/\Delta t}+\mathcal{O}(\Delta t).
\end{align}
Furthermore, considering the efficiency, the implementation of three-body terms can be accomplished layer by layer, wherein each layer only contains three-body operators with no overlap with each other, so that they can be applied to the MPS parallelly. We emphasize that since the Trotter error is of order $\Delta t$, the time interval $\Delta t$ should be small enough to avoid large TEBD error. The implementation of $U_2(\Delta t)$ is showed in Fig.~\ref{MPO}\textbf{b}.

\section{Experimental details}
\label{app:exp}
\subsection{Quantum circuit ansatz}

\begin{minipage}[t][8cm][t]{0.48\textwidth}
    \begin{algorithm}[H]
        \caption{Neuroevolution Method\label{alg:neuro}}
        \SetAlgoLined
        \KwOut{Quantum circuit ansatz approximating target unitary.}
        \KwIn{Elementary gate set $\mathcal{S}$, evolution unitary $U_2(\Delta t)$ and threshold $\beta$.}
        $\mathcal{G}$ = Direct\_Graph($\mathcal{S}$)\; 
        $\mathcal{C}$ = Random\_Generation\_of\_Quantum\_Circuit($\mathcal{G}$)\;
        $\mathcal{L}$ = Optimization($\mathcal{C}$, $U_2(\Delta t)$)\;
        \While{ $\min\{\mathcal{L}\}>\beta$}{
            $\quad\mathcal{C}$ = Quantum\_Circuit\_Extension($\mathcal{C}$, $\mathcal{G}$)\;
            $\quad\mathcal{L}$ = Optimization($\mathcal{C}$, $U_2(\Delta t)$)\;
        }
        \KwRet{\emph{arg}$\min_\mathcal{C}\{\mathcal{L}\}$}\;
    \end{algorithm}
\end{minipage}
\begin{minipage}[t][8cm][t]{0.48\textwidth}
    \begin{algorithm}[H]
        \caption{Optimization for a quantum circuit\label{alg:opt}}
        \SetAlgoLined
        \KwOut{Optimal parameters of the given quantum circuit.}
        \KwIn{A quantum circuit $C$, evolution unitary $U_2(\Delta t)$ and learning rate $\alpha$.}
        Randomly initialize $\bm{\theta}$\;
        $U_{\text{circuit}}(\bm{\theta})$ = Unitary($C$, $\bm{\theta}$)\;
        $L$ = $1-\text{Tr}\left[U_2(\Delta t)^\dagger U_{\text{circuit}}(\bm{\theta})\right]/d$\;
        \While{ $L>0.001$}{
            $\quad\bm{\theta}$ = $\bm{\theta} - \alpha\nabla_{\bm{\theta}}L$\;
            $\quad U_{\text{circuit}}(\bm{\theta})$ = Unitary($C$, $\bm{\theta}$)\;
            $\quad L$ = $1-\text{Tr}\left[U_2(\Delta t)^\dagger U_{\text{circuit}}(\bm{\theta})\right]/d$\;
        }
        \KwRet{$\bm{\theta}$}\;
    \end{algorithm}
\end{minipage}

\begin{figure}[t]
    \includegraphics[width=0.48\textwidth]{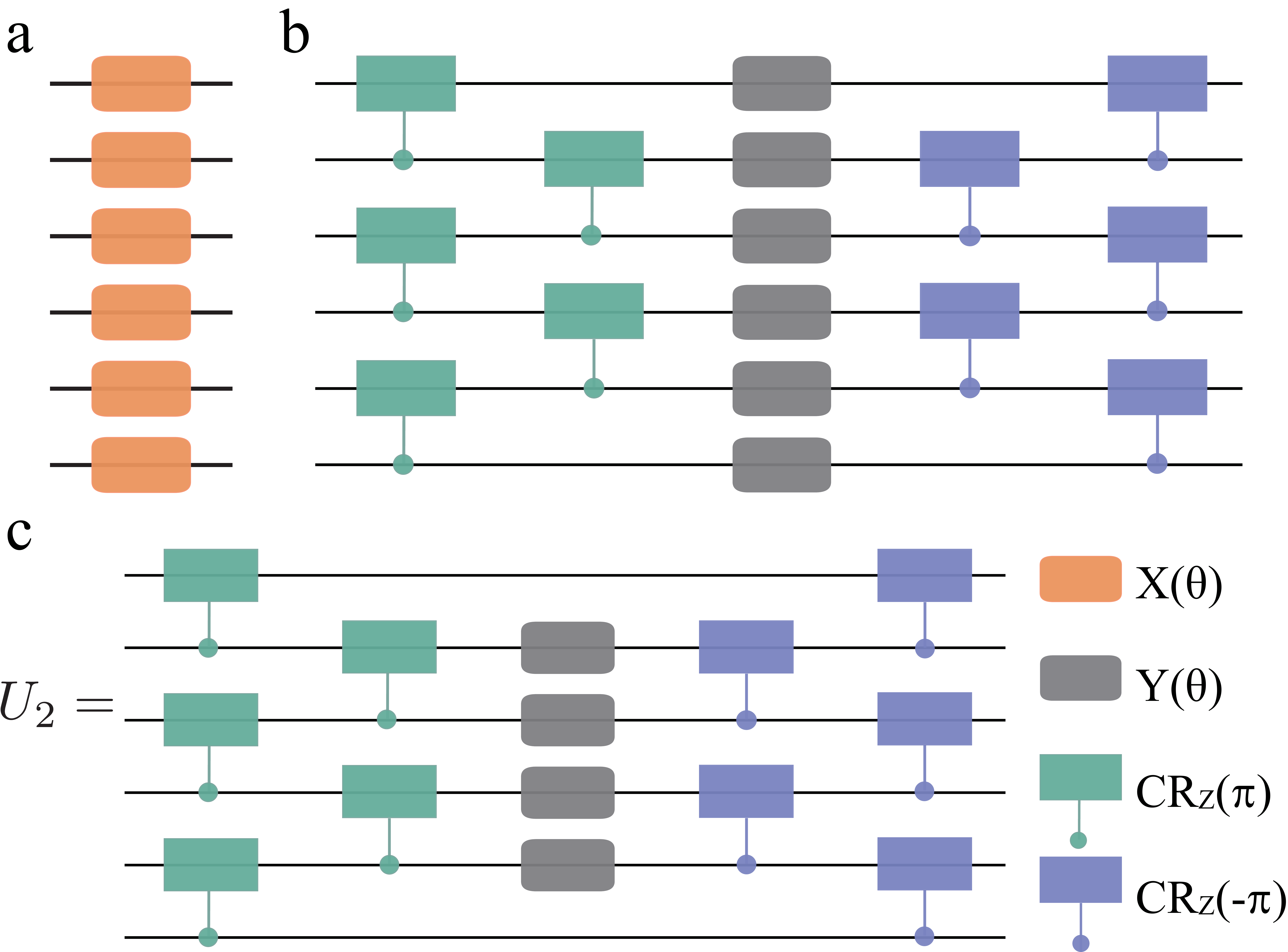}
    \caption{Quantum circuit ansatz used in experiments. \textbf{a.} The circuit ansatz for the time evolution unitary in the first time interval. \textbf{b.} The circuit ansatz for the time evolution unitary in the second time interval, where the system is in deep topological phase. \textbf{c.} The circuit ansatz for the time evolution unitary in the second time interval, where the system contains no two-body operators and no one-body operators: $U_2=\exp{(i\sum_kJ_k\hat\sigma^z_{k-1}\hat\sigma^x_{k}\hat\sigma^z_{k+1})}$.\label{circuit}}
\end{figure}

To observe the SPT time crystal on a real superconducting quantum computer, we need to decompose the time evolution unitary into a quantum circuit consisting of a series of experimentally implementable quantum gates. Due to the direct product form of the evolution unitary in the first time interval $U_1(t)=e^{-itH_1}$, this unitary can be represented as a quantum circuit using a layer of rotation gates along $x$ axis. Thus, it can be constructed and implemented relatively easily. As for the second time interval, the interaction among different sites leads the time evolution unitary far away from a direct product form, making things a little different.

With the progress of the research on the variational quantum circuits, we are able to adapt this method to construct the quantum circuit of the second time evolution unitary in our time crystal model. Variational quantum circuit is a powerful tool and is intensively investigated in recent years. Algorithms based on variational quantum circuits are of great potential in the noisy intermediate scale quantum (NISQ) era. There are many algorithms based on variational parameterized quantum circuits, such as variational quantum eigen-solver \cite{Peruzzo2014Variational}, quantum neural network \cite{Beer2020Training}, etc. They are proposed to solve practical problems. The major distinction between usually used quantum circuits and variational quantum circuits is that the gates contained in a variational quantum circuit are not fixed. They can be modified by switching their parameters using different parameter-updating algorithms. With the updating of those parameters, the unitary implemented by the variational circuit is varying. We end this updating procedure until a satisfactory result is obtained. 

Here, our target is to find a variational quantum circuit with some fine-tuned parameters nearly equivalent to the evolution unitary in the second time interval $U_2(t)=e^{-itH_2}$. We accomplish this target within two steps: finding an implementable variational quantum circuit ansatz which can be used to represent the target unitary, and updating those parameters contained in this circuit ansatz according to the evolution unitary of a particular random realization of the hamiltonian. 

We use the neuroevolution method \cite{Lu2020Markovian} to find a suitable variational circuit architecture. The elementary gates used in our experiments are single qubit rotation gates ($X,\ Y,\ Z$) and control-rotation gate along $z$ axis. And each of them contains a variational parameter. They can form various quantum circuit layers. i.e. quantum circuits with depth equal to one. We construct the direct graph of those layers consisting of those elementary gates with respect to the rules in \cite{Lu2020Markovian}. Then, a quantum circuit can be represented as a path in this graph. The later problem can be solved with the following procedures: 1) Randomly generate several variational quantum circuits with fixed length based on the directed graph; 2) Update parameters contained in those quantum circuits using gradient based algorithm to minimize the loss function $L(\bm{\theta})=1-\text{Tr}\left[U_2(\Delta t)^\dagger U_{\text{circuit}}(\bm{\theta})\right]/d$, where $U_2(\Delta t)$ is the evolution unitary in the second time interval, $U_{\text{circuit}}(\bm{\theta})$ is the unitary represented by the current quantum circuit with variational parameters $\bm{\theta}$, and $d$ is the dimensionality of the corresponding Hilbert space; 3) Chose quantum circuits with small loss values and extend them based on the direct graph to generate new circuits; 4) Iterate processes 2) and 3) until the loss is lower than a fixed threshold. The circuit ansatz giving the smallest loss value is regarded as the optimal ansatz representing the evolution unitary and is adapted in our experiments. We show the pseudo-code of this algorithm in Algorithm~\ref{alg:neuro}.

The quantum circuit ansatzs used in our experiments are showing in Fig. \ref{circuit}. We notice that the quantum circuit for the evolution unitary in the second time interval has a sandwich form $U_2(\Delta t)\approx WD(\bm{\theta})W^\dagger$, where $D(\bm{\theta})$ is a layer of single qubit rotation gates with $\bm{\theta}$ being the evolution-time dependent parameters. So for one driving period, the depth of the corresponding quantum circuit is fixed, i.e. for $n$ driving periods, the depth of the corresponding quantum circuit is $6n$, which is independent with the number of time slices in a period.

With this circuit ansatz in hand, we can then use it to construct the experiment circuits. For a particular disorder realization of the second hamiltonian deep in the topological phase ($J_k\gg V_k,h_k$), we begin with this ansatz containing randomly generated variational parameters $\bm{\theta}$. From this, the gradients of the loss function $L(\bm{\theta})$ with respect to those variational parameters are computed, and are used to update the current parameters $\bm{\theta}^{(n+1)}=\bm{\theta}^{(n)}-\alpha\nabla_{\bm{\theta}^{(n)}}L$, where $\alpha$ is the given learning rate (we usually chose $0.001\leq\alpha\leq0.01$). In our calculation, we iterate this optimization procedure until the operator fidelity \cite{Wang2009Operator} $\operatorname{Tr}\left[U_{2}(\Delta t)^{\dagger} U_{\text {circuit }}(\boldsymbol{\theta})\right] / d\geq0.999$ ($L(\bm{\theta})\leq 0.001$). Then, we take the quantum circuit with the final parameters as the approximation of the evolution unitary $U_2(\Delta t)$ in our experiments. We show the pseudo-code of this algorithm in Algorithm~\ref{alg:opt}.

We emphasize that this optimization procedure is suitable for small systems. Nevertheless, with the exponential growth of the dimensionality of the Hilbert space, the optimization for large systems is unpractical. It is helpful that the quantum circuit ansatz found by the neuroevolution method can exactly represent the evolution unitary $U_2(t)=e^{-itH_2}$, when $H_2$ consists no two-body operators and no one-body operators (as showing in Fig. \ref{circuit}\textbf{c}). This indicates that we can theoretically construct the corresponding quantum circuits for arbitrary many qubits. In our simulations and experiments, for systems of $L~\leq~8$, the two-body terms and one-body terms are considered and the parameters in the corresponding quantum circuits are obtained using gradient-based optimization method. For 14-qubit systems, we only consider the stabilizer terms and exactly construct those corresponding quantum circuits.

\subsection{Device overview and measurement setup}
To illustrate the idea of SPT time crystal, we select a chain of up to $L=14$ qubits in a superconducting quantum processor, which is a flip-chip device hosting an array of $6 \times 6$ qubits distributed in a square lattice. To realize high-fidelity controlled-Z (CZ) gates, we adopt the tunable coupler architecture~\cite{Yan2018Tunable} to mediated nearest neighbor qubit-qubit interactions, i.e., individual couplers are inserted between neighboring qubits with the qubit-coupler coupling strengths designed to be around 130 \text{MHz} for qubits at 6.5 \text{GHz}. All qubits (couplers) are of transmon type, with anharmonicities around 250 (350)~MHz and maximum resonance frequencies around 7 (10.5) \text{GHz}. Each qubit has its own control line, which takes microwave (XY) inputs for rotating the qubit state around $x$- or $y$-axis and flux bias (Z) pulses for tuning the qubit frequency and rotating the qubit state around $z$-axis; each coupler is frequency tunable via its own flux bias (Z) line, which guarantees that the effective coupling strength between two neighboring qubits at 6.5 \text{GHz} can be dynamically turned on, up to $-25$~\text{MHz}, or off, $\leq 0.25$~\text{MHz}.
Each qubit capacitively couples to its own readout resonator, designed in the frequency range from 4.1 to 4.4~\text{GHz}, for qubit state measurement. 9 readout resonators share one readout transmission line (TL) running across the processor chip, and 4 readout TLs can cover all 36 qubits in the processor.

\begin{figure*}[tbp]
\centering
\includegraphics[width=1\linewidth]{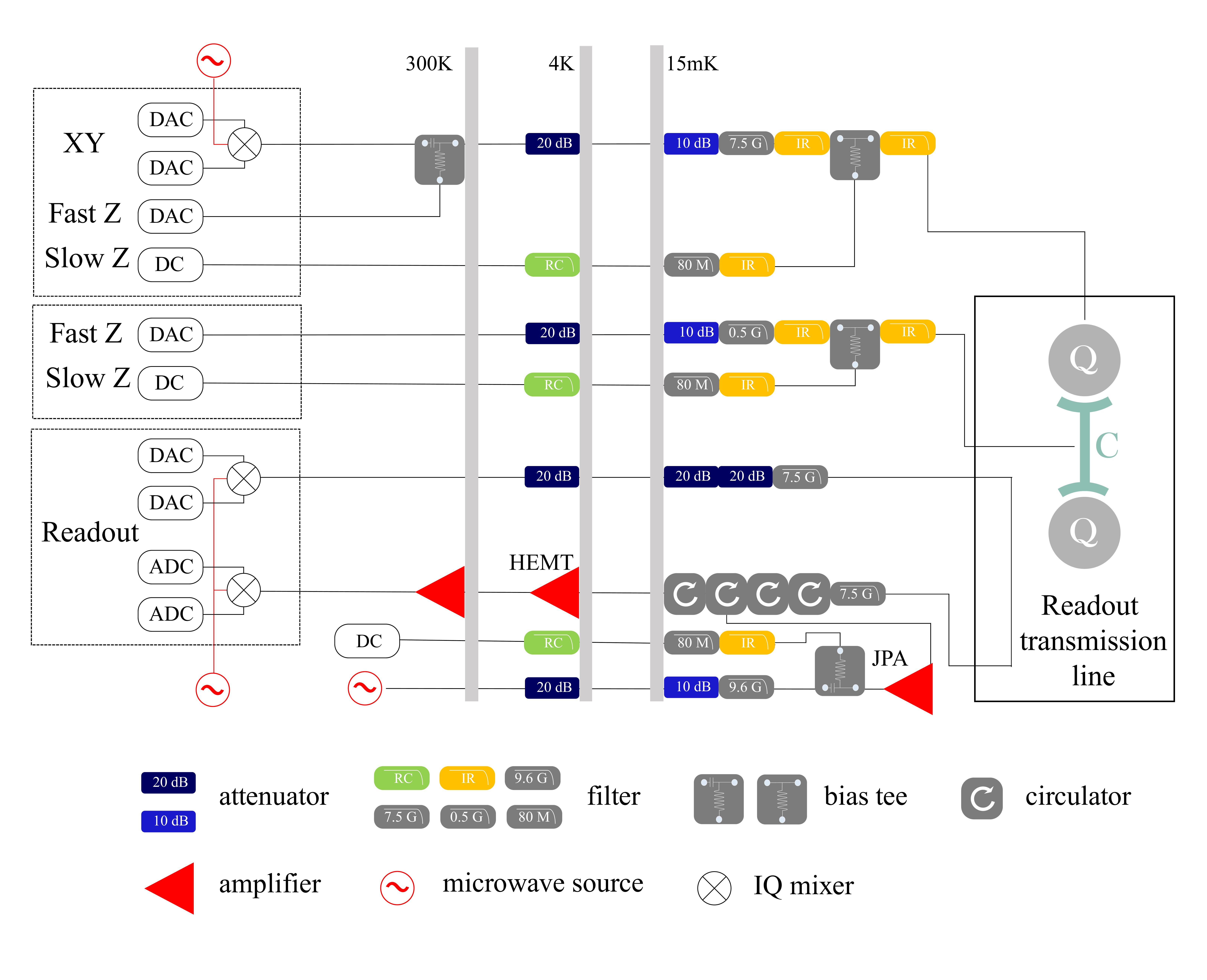}
\caption{Electronics and wiring setup illustrating how to synthesize and transmit the control/readout signals. Each qubit has three control channels: XY (microwave), fast Z (flux), and slow Z (flux). Each coupler has two control channels: fast Z and slow Z. Readout pulses are generated similarly as the XY signals, which are passed through the processor via the readout TLs.
All control and readout lines are well attenuated and filtered for noise shielding and delicate controlling.}
\label{fig:wiring}
\end{figure*}

The processor was fabricated using the flip-chip recipe: all qubits and couplers are located on sapphire substrate (top chip); most of the control/readout lines and readout resonators are located on silicon substrate (bottom chip). These two chips have lithographically defined base wirings, junction loops, and airbridges made of aluminum, and are galvanically connected via indium bumps with titanium under-bump metallization, as described elsewhere~\cite{zhang2021Synthesizing}. The indium bumps were formed by the lift-off method with 9 \text{$\mu m$}-thick indium deposited on both chips, after which these two chips were aligned and bonded together at room temperature to complete the flip-chip device. The indium bumps in our processor are not only for ground connectivity, but also for passing through control signals from the bottom chip to the top chip where the qubits are located.

The processor was loaded into a multilayer printed circuit board (PCB) enclosure, which was then mounted inside a dilution refrigerator (DR) with the base temperature down to 15 \text{mK}. Figure~\ref{fig:wiring} shows the schematics of the control/readout electronics and wiring setup. In this setup, the XY microwave signals and fast Z pulses synthesized by digital-to-analog converters (DACs) are first joined together at room temperature, which are attenuated and filtered at multiple cold stages of DR, and later combined with the slow Z (DC) pulses via homemade bias-tees at the mixing chamber stage of DR before being transmitted into the qubit control lines. The multiplexed readout signals are also heavily attenuated and filtered before going into the readout TLs of the processor to retrieve the qubit state information. To boost the signal-to-noise ratio (SNR), output signals from TLs are sequentially amplified by Josephson parametric amplifier (JPA), high electron mobility transistor (HEMT) amplifier, and room temperature (RT) amplifiers before being demodulated by analog-to-digital converters (ADCs) with 10-bit vertical resolution and 1.0~GS/s sampling rate. All arbitrary microwave signals are generated by mixing the DAC outputs with continuous microwaves using IQ mixers. DACs used to synthesize XY microwave signals and fast Z pulses in this experiment has {14}-bit vertical resolution and 300 \text{MHz} output bandwidth. Slow Z (DC) pulses are generated by commercial 16-bit DACs with maximum outputs of $\pm 2.5$~\text{V}.

\subsection{Single- and Two-qubit gates}
Single-qubit gates used in this experiment include X($\theta$), Y($\theta$), and Z($\theta$), which rotate the qubit state by an arbitrary angle $\theta$ around $x$-, $y$-, and $z$-axis, respectively. We realize X($\theta$) and Y($\theta$) by controlling the amplitude and phase of XY microwave pulses, and implement Z($\theta$) via the virtual Z gate~\cite{Mckay2017VZ}. Single-qubit gate errors are characterized by simulatneous randomized benchmarkings, yielding an average gate fidelity above 0.99 (see Tab.~\ref{tab:deviceParameters}).

\begin{table*}[tbp]
\caption{Device parameters calibrated during the experiment. $\omega_{j}^{\text{0}}$ is the maximum frequency of $Q_{j}$ at zero flux bias. $\omega_{j}$ is the idle frequency where we initialize $Q_{j}$ in $\ket{0}$ and subsequently apply single-qubit gates. $\eta_{j}$ is $Q_{j}$'s anharmonicity, which is approximately a constant within the frequency range relevant to this experiment. $\left[\omega_{j}^{\text{A (B)}}, \omega_{j+1}^{\text{A (B)}}\right]$ list the pairwise frequency values for neighboring two qubits where $\ket{11}$ and $\ket{02}$ in the two-qubit subspace are in near resonance for a CZ gate; the CZ gates for qubit pairs in the same group A (B) are implemented simultaneously when executing the multilayer quantum circuit to simulate SPT time crystal. $\omega_{j}^{\text{m}}$ is the readout frequency of $Q_{j}$ where we apply readout pulses to excite $Q_j$'s readout resonator for quantum state measurement. $\omega_{j}^{\text{r}}$ is the resonant frequency of $Q_j$'s readout resonator. $T_{1, j}$ and $T_{2, j}^*$ are the energy relaxation time and Ramsey dephasing time of $Q_j$, respectively. $F_{0, j}$ and $F_{1, j}$ are the readout fidelity values for $Q_{j}$ prepared in $\ket{0}$ and $\ket{1}$, respectively, which are used to correct raw probabilities to eliminate readout errors as done previously~\cite{Zhen2020Superrdiant}.
$e_{\text{sq}}$ list the single-qubit gate errors obtained by simultaneous randomized benchmarkings. $e^{\text{A} (\text{B})}_{\text{CZ}}$ lists the CZ gate errors obtained by both individual and simultaneous randomized benchmarkings for qubit pairs in group A (B). We note that the qubit parameters may slowly drift over time~\cite{Klimov2018T1, Ren2020Simultaneous}.}
\begin{center}
\begin{tabular}{p{3cm}<{\centering}p{0.9cm}<{\centering}p{0.9cm}<{\centering}p{0.9cm}<{\centering}p{0.9cm}<{\centering}p{0.9cm}<{\centering}p{0.9cm}<{\centering}p{0.9cm}<{\centering}p{0.9cm}<{\centering}p{0.9cm}<{\centering}p{0.9cm}<{\centering}p{0.9cm}<{\centering}p{0.9cm}<{\centering}p{0.9cm}<{\centering}p{0.9cm}<{\centering}}
  \hline
  \hline
                   Qubit                        & $Q_{1}$   & $Q_{2}$   & $Q_{3}$   & $Q_{4}$    & $Q_{5}$   & $Q_{6}$   & $Q_{7}$   & $Q_{8}$ & $Q_{9}$   & $Q_{10}$   & $Q_{11}$   & $Q_{12}$ & $Q_{13}$ & $Q_{14}$\\
  \hline
  $\omega_{j}^{\text{0}}/2\pi$ (\text{GHz})  &  7.021     & 6.970     &  7.000     & 6.864     & 6.840      & 7.028      & 6.819    & 6.879     & 6.770      & 6.854    & 6.818     & 6.962      & 6.925      & 6.970   \\

  $\omega_{j}/2\pi$ (\text{GHz}) &  6.450     & 6.730     & 6.890    & 6.651     & 6.565      & 6.750      & 6.676    & 6.600     & 6.520      & 6.620    & 6.721     & 6.893      & 6.838      & 6.960   \\
  
  $\eta_{j}/2\pi$ (\text{GHz})                 & 0.230    & 0.248      & 0.248    & 0.242     & 0.255      & 0.239      & 0.288    & 0.247     & 0.251      & 0.246    & 0.241     & 0.250      & 0.247      & 0.252   \\
  
  $\left[\omega_{j}^{\text{A}}, \omega_{j+1}^{\text{A}}\right]/2\pi$ (\text{GHz})     &\mcol{\cfill 6.414, 6.656}&\mcol{\cfill 6.893, 6.651}&\mcol{\cfill 6.275, 6.516}&\mcol{\cfill 6.632, 6.868}&\mcol{\cfill 6.349, 6.585}&\mcol{\cfill 6.717, 6.957}&\mcol{\cfill 6.684, 6.920}   \\
  
  $\left[\omega_{j}^{\text{B}}, \omega_{j+1}^{\text{B}}\right]/2\pi$ (\text{GHz})     &&\mcol{\cfillo 6.667, 6.898}&\mcol{\cfillo 6.651, 6.412}&\mcol{\cfillo 6.894, 6.657}&\mcol{\cfillo 6.766, 6.528}&\mcol{\cfillo 6.485, 6.722}&\mcol{\cfillo 6.910, 6.676}&    \\
 
  $\omega_{j}^{\text{m}}/2\pi$ (\text{GHz})  & 6.110      & 6.198  &  5.608      & 6.651    & 5.552     & 6.309   & 6.722     & 5.997     & 5.812    & 5.828     & 6.323      & 5.736      & 6.181      &6.423\\
  
  $\omega_{j}^{\text{r}}/2\pi$ (\text{GHz}) & 4.357  & 4.194  & 4.119      & 4.200    & 4.097     & 4.343   & 4.323     & 4.223     & 4.262    & 4.206     & 4.152      & 4.269      & 4.182      & 4.402\\
   
  $T_{1, j}$ (\text{$\mu$s})                  & 25       & 22     & 28        & 36         & 11       & 27        & 27      & 30        & 22        & 33        & 25       & 37         & 13        & 29    \\

  $T_{2, j}^*$ (\text{$\mu$s})                & 1.0       & 1.7       & 4.5       & 2.5        & 3.8       & 2.2      & 1.2     & 1.6       & 0.8       & 2.1       & 3.1      & 2.8        & 5.8      & 14.0      \\
  \hline
  $F_{0, j}$                                  & 0.950     & 0.955   & 0.945    & 0.888    & 0.951      & 0.951     & 0.961    & 0.956    & 0.868     & 0.880     & 0.959    & 0.935      & 0.980    & 0.970 \\
  
  $F_{1, j}$                                  & 0.876     & 0.862    & 0.834     & 0.888    & 0.886      & 0.942     & 0.859    & 0.900    & 0.890     & 0.905     & 0.900    & 0.898      & 0.919    & 0.937 \\
  \hline
  $e_{\text{sq}}$ (\%)        & 0.49     & 0.45      & 1.26     & 0.72     & 0.38     & 0.69        & 0.66    & 0.47      & 0.84     & 0.60      & 0.33      & 0.55      & 0.55     & 0.45     \\
\hline

  $e^{\text{A}}_{\text{CZ}}$ (\%) (\text{Indiv.})
  &\mcol{\cfill 1.06}&\mcol{\cfill 0.22}&\mcol{\cfill 1.79}&\mcol{\cfill 0.74}&\mcol{\cfill 0.99}&\mcol{\cfill 0.37}&\mcol{\cfill 1.09}   \\
    
  $e^{\text{B}}_{\text{CZ}}$ (\%) (\text{Indiv.})
  & &\mcol{\cfillo 0.29}&\mcol{\cfillo 1.24}&\mcol{\cfillo 0.59}&\mcol{\cfillo 1.77}&\mcol{\cfillo 0.78}&\mcol{\cfillo 1.68} \\
  \hline
  $e^{\text{A}}_{\text{CZ}}$ (\%) (\text{Simu.})
  &\mcol{\cfill 3.46}&\mcol{\cfill 0.99}&\mcol{\cfill 3.00}&\mcol{\cfill 0.76}&\mcol{\cfill 2.03}&\mcol{\cfill 0.79}&\mcol{\cfill 1.33}   \\
  $e^{\text{B}}_{\text{CZ}}$ (\%) (\text{Simu.})
  &&\mcol{\cfillo 0.76}&\mcol{\cfillo 0.51}&\mcol{\cfillo 0.81}&\mcol{\cfillo 2.29}&\mcol{\cfillo 0.97}&\mcol{\cfillo 2.05} \\
\hline
\hline
\end{tabular}
\end{center}
\label{tab:deviceParameters}
\end{table*}

\begin{figure*}[tbp]
\centering
\includegraphics[width=1.0\linewidth]{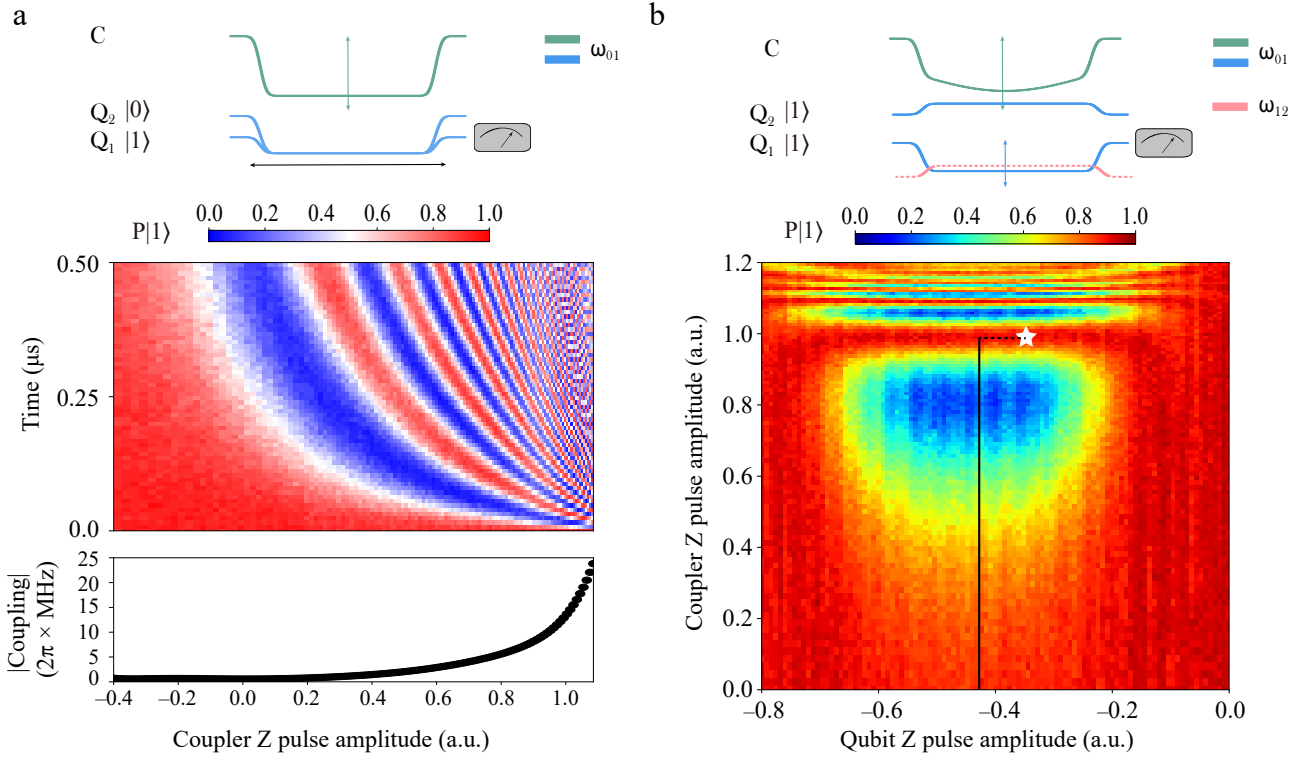}
\caption{Two-qubit CZ gate. a, two-qubit swap dynamics showing the dynamical range of the effective coupling strength as tuned by the coupler. b, $|1\rangle$-state population landscape for $Q_1$ resulting from the $\ket{11}$ and $\ket{02}$ interaction after initializing $Q_1$-$Q_2$ in $|11\rangle$. The white star marks the vicinity of the gate parameters used for CZ and lines indicate how we sweep Z parameters to approach this vicinity.}
\label{fig:twoQubitGate}
\end{figure*}

The basic structure to implement the CZ gate consists of two flux tunable qubits and one flux tunable coupler, which are, respectively, noted as $Q_1$, $Q_2$, and $C$ here for the clarity of description. The effective coupling strength is composed of a direct coupling strength between two qubits and a part mediated by the coupler, which can be continuously adjusted by controlling the flux or frequency of the coupler. The Hamiltonian of this three body system is written as
\begin{equation}
\begin{aligned}
H / \hbar &=\sum_{i=1,2, c} \omega_{i} a_{i}^{\dagger} a_{i}+\frac{\alpha_{i}}{2} a_{i}^{\dagger} a_{i}^{\dagger} a_{i} a_{i} \\
&+\sum_{i < j} g_{i j}\left(a_{i} - a_{i}^{\dagger}\right)\left(a_{j}- a_{j}^{\dagger}\right),
\end{aligned}
\end{equation}\\
The effective coupling strength between qubits is
\begin{equation}
    \Tilde{g} = g_{1c}g_{2c}\left(\frac{\omega_{1}}{\omega_{c}^2-\omega_{1}^2}+ \frac{\omega_{2}}{\omega_{c}^2-\omega_{2}^2}\right)+g_{12}.
\end{equation}
In Fig.~\ref{fig:twoQubitGate}\textbf{a} we plot the dynamic range of $\Tilde{g}$ (bottom panel) processed using the two-qubit swap dynamics after initializing $Q_1$-$Q_2$ in $|10\rangle$, which shows that the effective coupling strength is tunable in the range from $-25$~MHz to $\le 0.25$~MHz.
Experimentally we can apply single-qubit gates while locating the coupler at around 10.5~GHz to turn off $\Tilde{g}$.

To realize the CZ gate, we apply a flux bias (fast Z) pulse to steer the coupler's frequency in a trajectory of 10.5$\rightarrow$7.3$\rightarrow$10.5~GHz, and meanwhile we turn on the fast Z pulses to bring $Q_1$ and $Q_2$ from their idle frequencies to the pair of values in $\left[\omega_{j}^{\mathrm{A}(\mathrm{B})}, \omega_{j+1}^{\mathrm{A}(\mathrm{B})}\right]$ depending on which pair we use, where $\ket{11}$ and $\ket{02}$ in the two-qubit subspace are in near resonance (see Tab. ~\ref{tab:deviceParameters}). After a finite period for this diabatic interaction, an unitary two-qubit gate equivalent to CZ up to trivial single-qubit phase factors can be obtained as
\begin{equation}
\left(\begin{array}{cccc}
1 & 0 & 0 & 0 \\
0 & e^{i \phi_{1}} & 0 & 0 \\
0 & 0 & e^{i \phi_{2}} & 0 \\
0 & 0 & 0 & e^{i \phi_{3}}
\end{array}\right).
\label{CZmatrix}
\end{equation}
A sine decorated square pulse with the amplitude $A=z_{0} \times\left[1-r+ r\sin \left(\pi \frac{t}{t_{\text{gate}}}\right)\right]$ is used for the coupler in order to minimize state leakages. Experimentally we fix $r=0.3$ and only fine-tune the parameter $z_{0}$. All pulses are digitally smoothed by convolving them via a Gaussian window with $\sigma=2$~ns before applying our pulse calibration routines~\cite{Chao2017GHZ}.
The CZ gate pulse duration is $30$ \text{ns}, and there are additional 5 ns padding times before and after the 30-ns gate in compensation for the finite small tails of the smoothed pulse.

Individual CZ gates are calibrated following the procedure below:
\begin{enumerate}
\item  Optimize coupler Z bias amplitude for minimum state leakages: We initialize $Q_1$-$Q_2$ in $|11\rangle$ and fix their frequency detuning at $\omega_1-\omega_2 \approx -2\pi\times 250$~MHz, following which we apply the sine decorated square pulse with a total length of 40 ns to the coupler. We search for the optimized pulse amplitude $z_0$ which maximizes the $\ket{1}$-state population for $Q_1$, i.e., minimum state leakages. In Fig.~\ref{fig:twoQubitGate}\textbf{b} we plot the whole landscape of state leakages as functions of the Z bias amplitudes of both the coupler and $Q_1$, where the black solid line indicates how we sweep the coupler Z pulse amplitude.
\item Optimize phase factors: We fix the coupler Z pulse and sweep $Q_1$ Z pulse amplitude using different initial states to calculate the three phase factors in Eq.~\ref{CZmatrix}, aiming that $\phi_{3}-\phi_{2}-\phi_{1} = \pi$. The black dash line in Fig.~\ref{fig:twoQubitGate}\textbf{b} shows the routine how we sweep the qubit Z pulse amplitude. We apply virtual Z gates to remove the trivial single-qubit phases.
\item Fine-tune gate parameters according to RB: We choose the RB sequence fidelity as a goal function to optimize relevant gate parameters, including the Z pulse amplitudes of both qubits and the coupler, and the single-qubit phases. We use the Nelder-Mead (NM) method to speed up the parameter optimization process.
\end{enumerate}

\begin{figure}
\includegraphics[width=0.49 \textwidth]{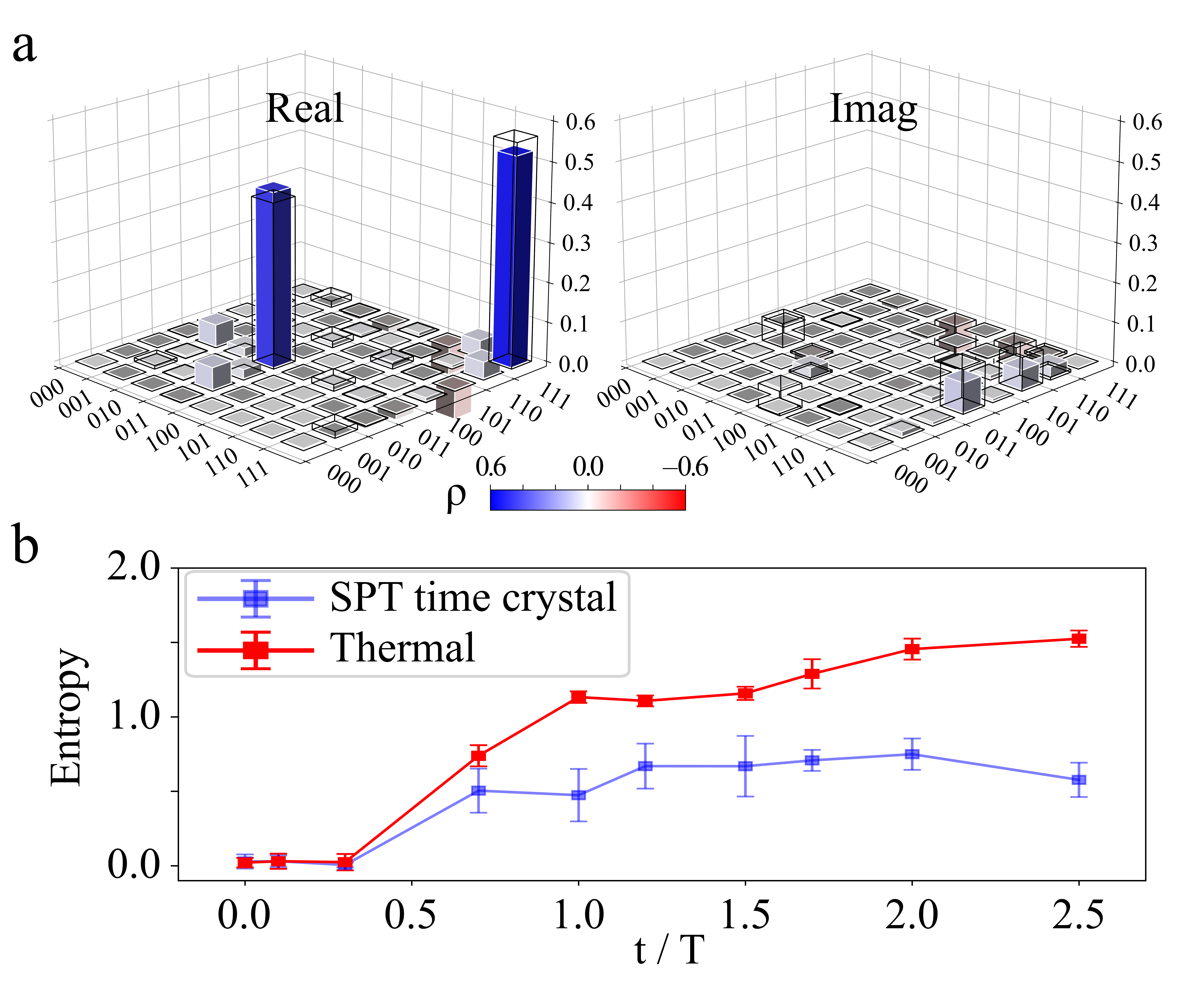}

\caption{\text{Entanglement dynamics.}  \textbf{a}, Tomography of the reduced density matrix for the second half of a six-qubit chain after one driving period of one instance. The real and imaginary parts are shown in the left and right panel, respectively. \textbf{b}, Entanglement growth in both the thermal and time crystal regions, averaging from 9 instances.  
\label{fig:Entropy}}
\end{figure}

\section{Dynamics of entanglement}

Unlike thermal phases without disorder or Anderson localized phases without interaction, where entanglement grows ballistically \cite{Cheneau2012Light,Richerme2014Non,Jurcevic2014quasiparticle} or saturates to an area law at long time,  the entanglement entropy of a MBL system grows logarithmically and saturates to an volume law in the long time limit \cite{Bardarson2012Unbounded}.  In our experiment, we also extract the entanglement  dynamics, through a full quantum state tomography of the reduced density state of half the system. In Fig.~\ref{fig:Entropy}\textbf{a}, we plot the reduced density matrix $\rho_{\text{half}}$ for a single random instance of the Hamiltonian at the end of one driving period. With the tomography data of the reduced density states at different times, we extract the desired information about the entanglement growth for a SPT time crystal. Our results are plotted in Fig. \ref{fig:Entropy}\textbf{c}. From this figure, it is clear that in the thermal region, the entanglement grows quickly and saturate to a maximal volume law ($\sim \frac{L}{2}\ln 2$) after about three driving period. In contrast, in the SPT time crystalline region, the entanglement grows much slower. Due to the decoherence and other imperfections in our experiment, we are not able to observe the logarithmic growth of the entanglement. This not only demands a significant improvement of the gate fidelity and a substantial increase of the coherence time, but also a more efficient and scalable approach to measure entanglement for a many-body system.

\end{document}